\documentclass[12pt,aps,pre,preprint,showkeys]{revtex4}

\usepackage{graphicx}
\usepackage{subfigure}
\usepackage{amsmath}
\usepackage{tikz}
\usepackage{setspace}
\usepackage{amssymb}
\usepackage{bm}
\usepackage{mathtools}

\DeclarePairedDelimiter\ceil{\lceil}{\rceil}

\begin{document}

\title{Phase diagrams of antiferromagnetic $XY$ model on a triangular lattice with higher-order interactions}
\author{M. Lach},
\author{M. \v{Z}ukovi\v{c}}
\email{milan.zukovic@upjs.sk}
\affiliation{Department of Theoretical Physics and Astrophysics, Institute of Physics, Faculty of Science, Pavol Jozef \v{S}af\'arik University in Ko\v{s}ice, Park Angelinum 9, 041 54 Ko\v{s}ice, Slovak Republic}
\date{\today}

\begin{abstract}
We study effects of higher-order antinematic interactions on the critical behavior of the antiferromagnetic (AFM) $XY$ model on a triangular lattice, using Monte Carlo simulations. The parameter $q$ of the generalized antinematic (ANq) interaction is found to have a pronounced effect on the phase diagram topology by inducing new quasi-long-range ordered phases due to competition with the conventional AFM interaction as well as geometrical frustration. For values of $q$ divisible by 3 the conflict between the two interactions results in a frustrated canted AFM phase appearing at low temperatures wedged between the AFM and ANq phases. For $q$ nondivisible by 3 with the increase of $q$ one can observe the evolution of the phase diagram topology featuring two ($q=2$), three ($q=4,5$) and four ($q \geq 7$) ordered phases. In addition to the two phases previously found for $q=2$, the first new phase with solely AFM ordering arises for $q=4$ in the limit of strong AFM coupling and higher temperatures by separating from the phase with the coexisting AFM and ANq orderings. For $q=7$ another phase with AFM ordering but multimodal spin distribution in each sublattice appears at intermediate temperatures. All these algebraic phases also display standard and generalized chiral long-range orderings, which decouple at higher temperatures in the regime of dominant ANq (AFM) interaction for $q \geq 4$ ($q \geq 7$) preserving only the generalized (standard) chiral ordering.
\end{abstract}




\maketitle

\section{Introduction}
\hspace*{5mm} Although the two-dimensional $XY$ model with short-range interactions and continuous symmetry lacks any form of true long-range ordering (LRO)~\cite{Mermin_Wagner_1966}, it still can undergo the Berezinskii-Kosterlitz-Thouless (BKT) phase transition~\cite{BKT1, BKT2}. The low-temperature phase remains critical for all temperatures below the transition point at $T_{BKT}$, displaying quasi-long-range ordering (QLRO) with an algebraically decaying spin-spin correlation function due to bound pairs of vortices and antivortices. At $T_{BKT}$, the infinite-order phase transition leads to unbinding of the vortex-antivortex pairs and a completely disordered phase with an exponentially decaying correlation function. The behavior of the standard $XY$ model is well understood, nevertheless, its many generalizations remain an active subject of study due to rich and interesting critical behavior~\cite{Domany_1984, Lee_Grinstein_1985, Korshunov_1985, Sluckin_1988, Carpenter_1989, Romano_2002, Hayden_2010, hlub_2008, QI_2013, Hubscher_2013, Park-nematic, qin_2009, dian_2011, Poderoso-2011, Canova-2014, Canova-2016, zuko-kala_2017, zuko-kala_2018, zuko_2019, Lach_Zukovic_2020}, in connection with experimental realizations~\cite{LQ_films, Superfluids1, Superfluids2, Taroni_2008, CuprateSuperconductors}, and potential for interdisciplinary applications~\cite{Grason_2008, Cairns2016,Clark2016,zukovic_frustrated_2016, zuko-hris_2018}. 

Most of the above generalizations of the standard $XY$ model are based on inclusion of higher-order interactions. Namely, besides the usual magnetic interaction with spin angle periodicity of $2\pi$, there is an additional (generalized) nematic interaction characterized by a positive integer $q$ such that its periodicity is $2\pi/q$. Such a term produces vortices with noninteger $1/q$ winding number, which compete with the conventional vortices and antivortices generated by the magnetic interaction. The Hamiltonian of such a generalized model can be expressed in the form
\begin{equation}
\label{Generalized_Hamiltonian}
\mathcal{H} = -J_1\sum_{\langle i,j\rangle} \cos(\phi_{i,j}) - J_q\sum_{\langle i,j\rangle} \cos( q\phi_{i,j}),
\end{equation}   
where $\phi_{i,j} = \phi_i - \phi_j$ is the angle between two neighboring spins at sites $i$ and $j$, and $J_1$ and $J_q$ are the exchange interaction parameters. The first term $J_1$ is a usual magnetic, i. e. FM ($J_1 > 0$)  or AFM ($J_1 < 0$) coupling, while the second term $J_q$ represents a generalized nematic, Nq ($J_q > 0$) or ANq ($J_q < 0$) interaction.

When $q = 2$ and both $J_1$ and $J_2$ are positive, the inclusion of the nematic N2 term leads to the appearance of the FM and the nematic N2 QLRO phases, with the phase transition belonging to the Ising universality class~\cite{Lee_Grinstein_1985, Korshunov_1985, Sluckin_1988, Carpenter_1989, QI_2013, Hubscher_2013,  Nui_2018}. Additionally, theoretical investigations of the model with the competing FM ($J_1>0$) and AN2 ($J_2<0$) interactions revealed the existence of a new canted ferromagnetic phase at very low temperatures wedged between the FM and AN2 phases~\cite{dian_2011, zuko_2019}. A recent series of papers~\cite{Poderoso-2011, Canova-2014, Canova-2016} studied the effect of the gradual increase of the parameter $q>2$ on the critical properties of the model with both $J_1$ and $J_q$ positive. It was found that the higher-order interactions lead to a qualitatively different phase diagrams than the one observed for $q=2$. In particular, for $q \geq 4$ they revealed up to two additional ordered phases originating from the competition between the FM and Nq couplings, with the phase transitions belonging to a variety of universality classes.

The above studies assumed a bipartite (square) lattice, on which the character of the magnetic interaction (the sign of $J_1$) is not expected to change the phase diagram. However, in the model on a nonbipartite (e.g., triangular) lattice the AFM ($J_1<0$) interaction leads to geometrical frustration, which can drastically change the critical behavior. Such a model has been intensively studied~\cite{Miyashita_1984,Lee_1986,Lee_1998,Korshunov_2002,Hasenbusch_2005,Obuchi_2012} due to the possibility of spin-chirality decoupling, where the transitions to the magnetic QLRO and the vector chiral LRO phases occur at different temperatures. The inclusion of the AN2 ($J_2<0$) term leads to the emergence of the AFM and AN2 phases, with the transition belonging to the Ising universality class~\cite{Park-nematic}. Thus, the phase diagram topology as well as the character of the phase transition between the ordered phases is similar to the FM-N2 case. Nevertheless, the AFM-AN2 model additionally displays a chiral LRO which slightly extends above the BKT line. 

In our recent study of the geometrically frustrated AFM-ANq model on a triangular lattice~\cite{Lach_Zukovic_2020} we have demonstrated that the nematic parameter increased to $q = 3$ induces a new peculiar canted antiferromagnetic (CAFM) phase. It appears at low temperatures, situated between the AFM and AN3 phases, with the AFM-CAFM and AN3-CAFM phase transitions belonging to the weak Ising and weak three-state Potts universality classes, respectively. Thus, compared to the nonfrustrated FM-Nq model in which the increasing $q>2$ first changed the phase diagram topology for $q=4$~\cite{Canova-2016}, in the frustrated AFM-ANq case such a change occurred already for $q = 3$. Apparently, the effect of the increasing $q$ in the two cases is different and, thus, we find it interesting to study the evolution of the phase diagram topology with the increasing order of the generalized nematic coupling also in the latter case. The groundwork for such a study has already been laid in our previous work~\cite{zukovic_frustrated_2016}, which focused on the ground states of this model for up to $q = 8$. In the present paper we extend the investigation to finite temperatures with the goal to establish phase diagrams of the AFM-ANq models on the triangular lattice for $4 \leq q \leq 15$.

\section{Simulations} 
We perform Monte Carlo (MC) simulations of the model (\ref{Generalized_Hamiltonian}) using the Metropolis algorithm. We consider the generalized nematic parameter $q = 4, 5, ..., 10$ (with checks up to $q=15$) and the interaction parameters $J_1$ and $ J_q$ in the form $J_1 = -\Delta$, $J_q = \Delta-1$, with $\Delta \in [0, 1]$ to cover the interactions between the pure ANq $(\Delta = 0)$ and the pure AFM $(\Delta = 1)$ limits. Periodic boundary conditions were used to simulate systems with a linear size $L$. Owing to the highly efficient parallelized implementation on graphical processing units (GPU) we were able to run extensive simulations of relatively large system sizes. For calculation of thermal averages of various quantities of interest we typically use $L = 384$ and for studying spin distributions we consider much larger sizes up to $L = 1536$. In order to obtain the thermal averages, at each temperature step $5\times10^6$ MC sweeps were performed, with $20\% $ discarded for equilibration. The simulation at the next temperature starts from the final configuration obtained at the previous temperature, which helps to keep the system near the equilibrium throughout the whole simulation.

To detect phase transitions between various phases and to determine the respective phase diagrams, we calculate the following quantities: the internal energy per spin
\begin{equation}
 e =\frac{\langle \mathcal{H} \rangle}{L^2},
 \end{equation}
 the specific heat per spin
\begin{equation}
c = \frac{\langle \mathcal{H}^2\rangle  -  \langle \mathcal{H}\rangle^2}{T^2 L^2},
\end{equation}
 the magnetic $(m_1)$ and generalized nematic $(m_2, m_3, ..., m_q)$ QLRO parameters
\begin{equation}
m_k = \frac{\langle M_k\rangle}{L^2} = \frac{1}{L^2} \left\langle  \sqrt{3 \sum_{\alpha = 1}^3 \textbf{M}^2_{k\alpha}} \right\rangle, k = 1, 2, ...,  q; \alpha = 1, 2, 3,
\end{equation}
where $\textbf{M}_{k\alpha}$ is the $\alpha$-th sublattice QLRO parameter vector given by
\begin{equation}
\textbf{M}_{k\alpha} =  \left(\sum_{i \in \alpha} \cos(k\phi_{\alpha i}),\sum_{i \in \alpha} \sin(k\phi_{\alpha i})\right),
\end{equation}   
and finally, the standard ($\kappa_1$) and generalized ($\kappa_2, \kappa_3, ..., \kappa_q$) staggered chiralities
\begin{equation}
\kappa_k = \frac{\langle K_k\rangle}{L^2}= \frac{1}{2L^2}\left\langle \left| \sum_{p^+\in\bigtriangleup} \kappa_{kp^+} -  \sum_{p^-\in\bigtriangledown} \kappa_{kp^-} \right| \right\rangle, k = 1, 2, ..., q,
\end{equation}
where $\kappa_{kp^+}$ and $\kappa_{kp^-}$ are the local generalized chiralities for each elementary plaquette of upward and downward triangles, respectively, defined by
\begin{equation}
\kappa_{kp} = 2\{\sin[k(\phi_2-\phi_1)] + \sin[k(\phi_3-\phi_2)]+ \sin[k(\phi_1-\phi_3)]\}/3\sqrt{3}.
\end{equation}
The susceptibilities of the respective order parameters can be defined as
\begin{equation}
\chi_o = \frac{1}{TL^2} (\langle \mathcal{O}^2\rangle  -  \langle \mathcal{O}\rangle^2), \mathcal{O} = M_1, M_2, ...,  M_q; K_1, K_2,..., K_q.
\end{equation}
The above quantities are useful in identifying the character of the QLRO (from the order parameters) as well as in rough determination of the phase boundaries (from the response functions). We note that the focus of the present study is the evolution of the phase diagram topology in a wide parameter space rather than precise determination of the phase boundaries. The latter would involve a finite-size scaling (FSS) analysis, which would require tremendous computational demands even on GPU and thus we leave such analysis for future considerations. 

Nevertheless, in the cases when the above quantities do not provide conclusive evidence, we further perform a correlation analysis to more reliably determine different phases. Such analysis is based on FSS of the QLRO parameters, obeying the scaling law
\begin{equation} 
\label{eq:fss_o_bkt}
m_k(L) \propto L^{-\eta_{m_k} (T)}, 
\end{equation}
where $\eta_{m_k}(T)$ is the temperature-dependent critical exponent of the correlation function $G_k=\langle \cos(k\phi_{i,j}) \rangle$, $k=1,\ldots,q$. The transition temperature from the phase characterized by the parameter $m_k$ can be determined as the temperature at which the critical exponent crosses to the value $\eta_{m_k}=1$, characteristic for an exponential decay of the correlation function $G_k$.
      
\section{Results}

The ground states of the present model were investigated in Ref.~\cite{zukovic_frustrated_2016}. It was concluded that for $q$ nondivisible by 3 the generalized nematic term prefers relative phase angles, which include $\Delta \phi = \pm 2\pi/3$, characteristic for the chiral AFM order observed in the standard $XY$ model without higher-order couplings. Thus, there is no conflict between the magnetic and generalized nematic interactions and the system shows the chiral AFM ordering. However, for $q$ divisible by 3 such sublattice-uniform ordering disappears. Instead, the neighboring spins belonging to different sublattices align forming phase angles with the values dependent on the ratio of the exchange interactions $J_1$ and $J_q$ such a way that on each triangular plaquette two neighbors are oriented almost parallel with respect to each other and
almost antiparallel with respect to the third one. Such a microscopic arrangement results in a macroscopic degeneracy, loss of the sublattice uniformity, and the canted AFM (CAFM) phase. 

Our recent study of the model with $q=3$~\cite{Lach_Zukovic_2020} showed that the CAFM phase extends to finite temperatures and crosses to the AFM phase for dominant $J_1$ or to the AN3 phase for dominant $J_3$ or straight to the paramagnetic phase for $J_1$ and $J_3$ of comparable strengths. Thus, the increase of $q$ from 2 to 3 resulted in the change of the phase diagram topology from the one with two phases (AFM and AN2) to that featuring three phases (AFM, AN3 and CAFM). The results of the present study indicate that such a topology remains unchanged for any $q$ divisible by 3 up to at least $q=15$. Therefore, in the following, we will focus on the study of the phase diagram topology evolution with the increasing $q>3$ for the values nondivisible by 3. As we will see, even though there is no conflict between the AFM and ANq interactions for such cases there is still competition present, which will result in the formation of new phases.

\begin{figure}[t!]
\centering
\subfigure{\includegraphics[scale=0.34,clip]{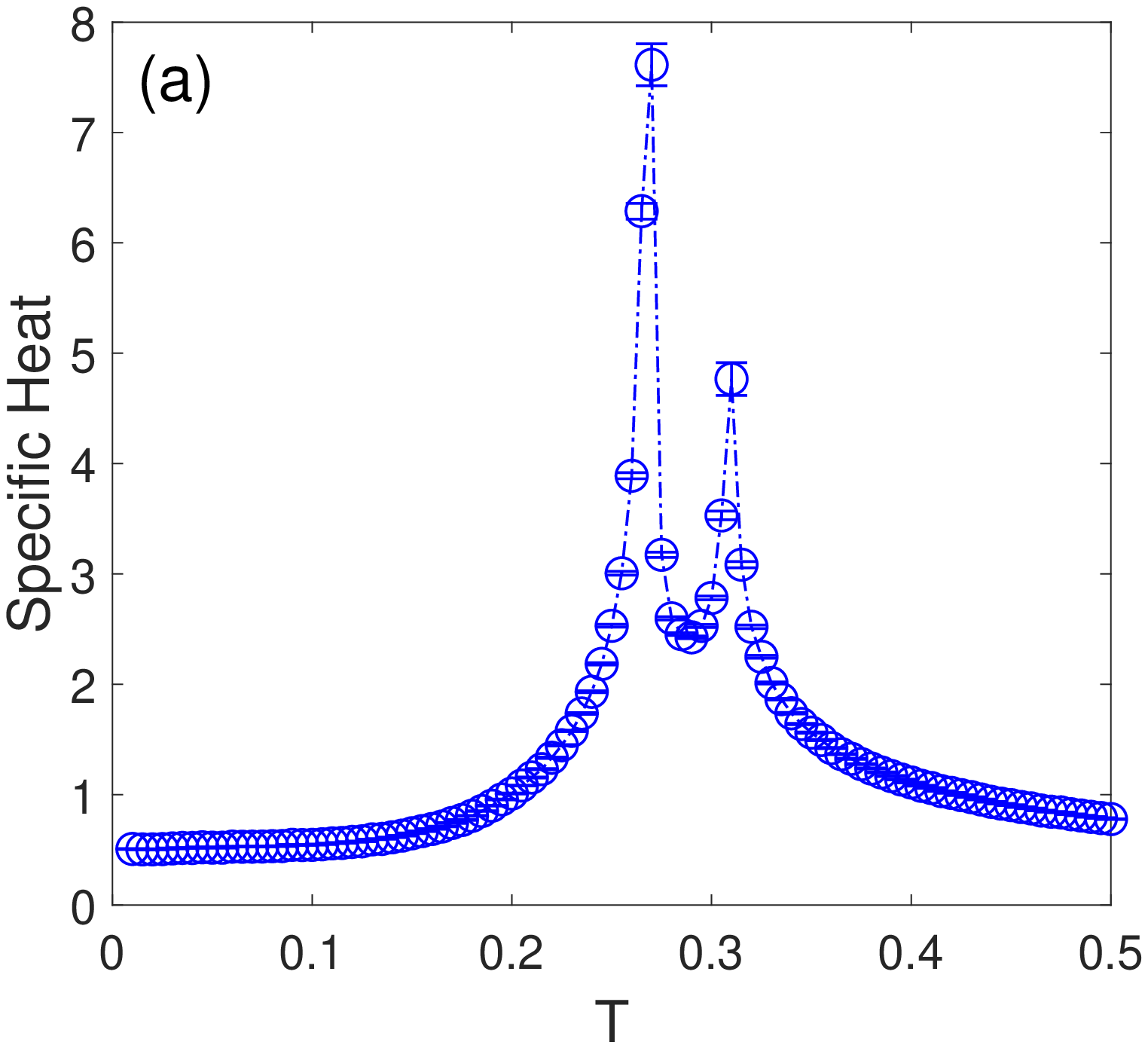}\label{fig:C_q4_x04}}
\subfigure{\includegraphics[scale=0.34,clip]{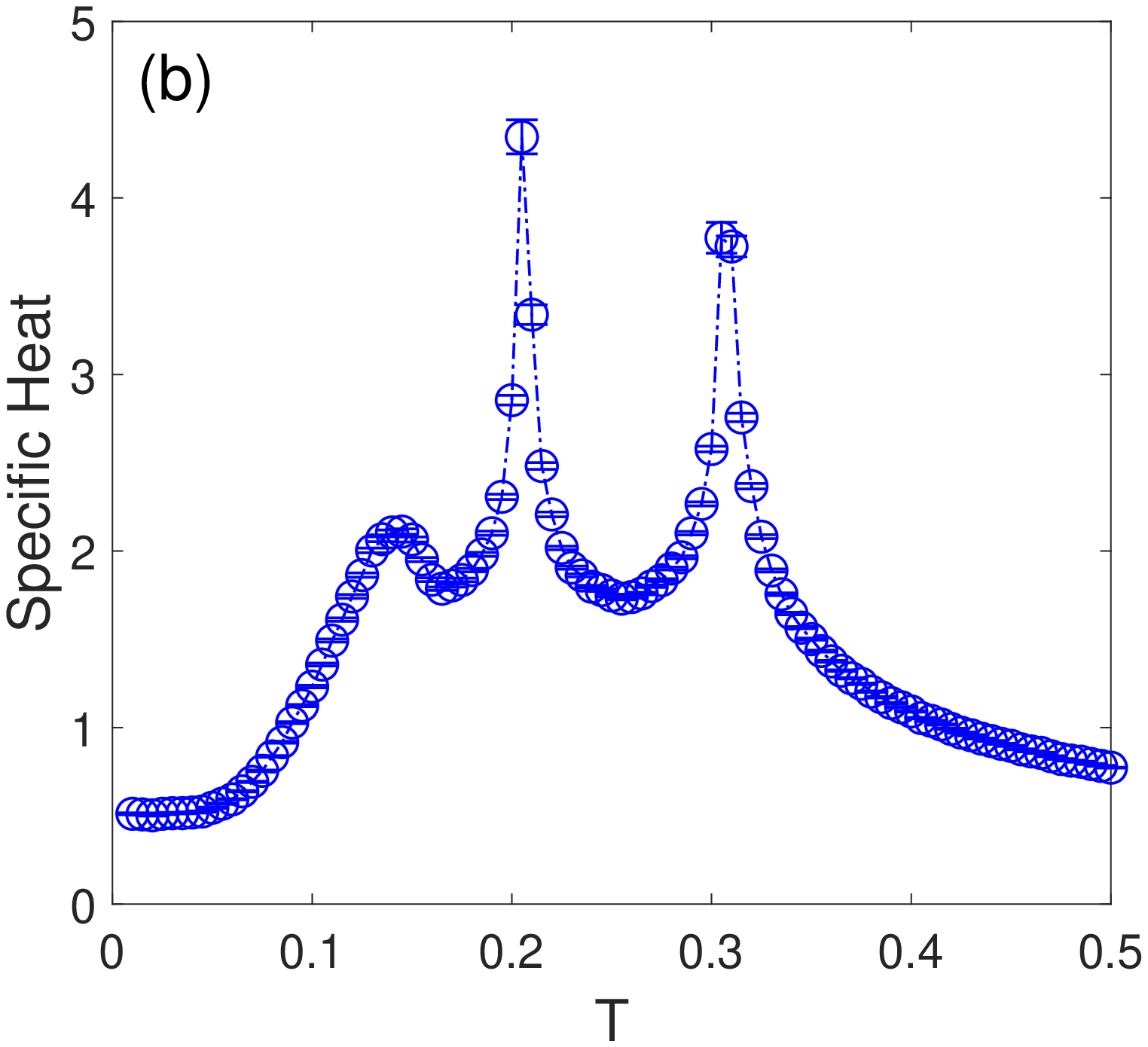}\label{fig:C_q7_x04}}
\subfigure{\includegraphics[scale=0.34,clip]{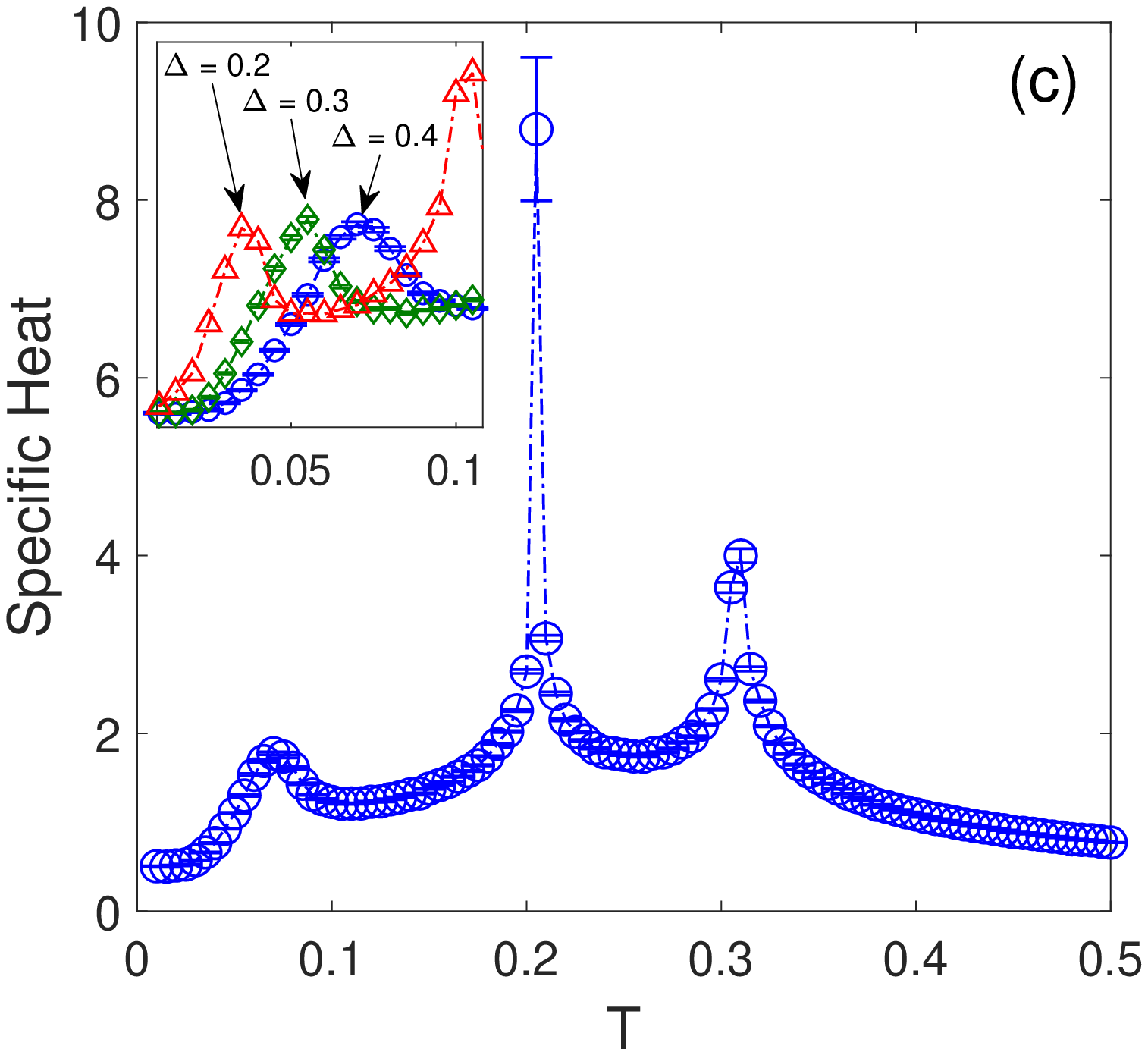}\label{fig:C_q10_x04}}\\
\subfigure{\includegraphics[scale=0.34,clip]{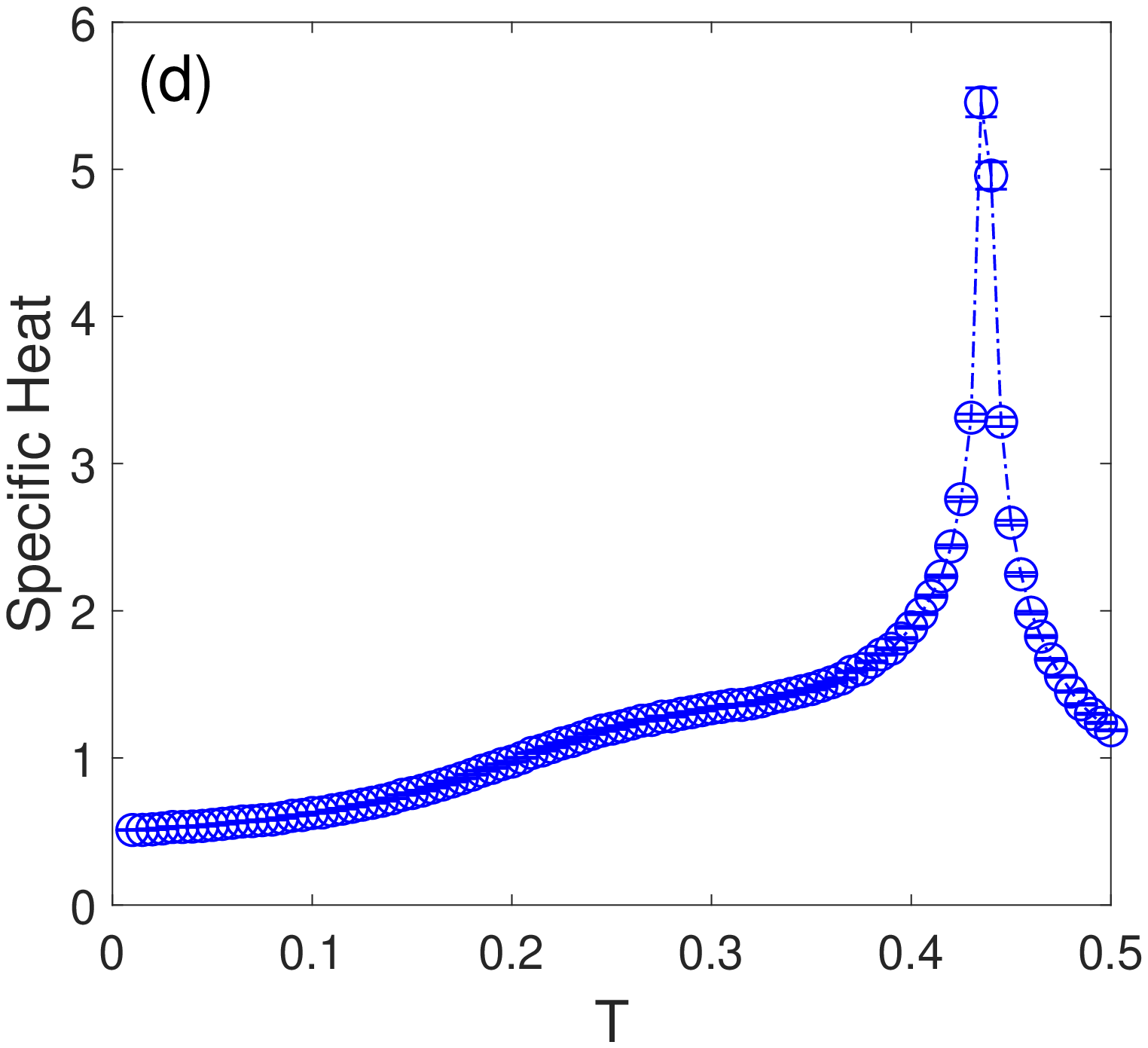}\label{fig:C_q4_x08}}
\subfigure{\includegraphics[scale=0.34,clip]{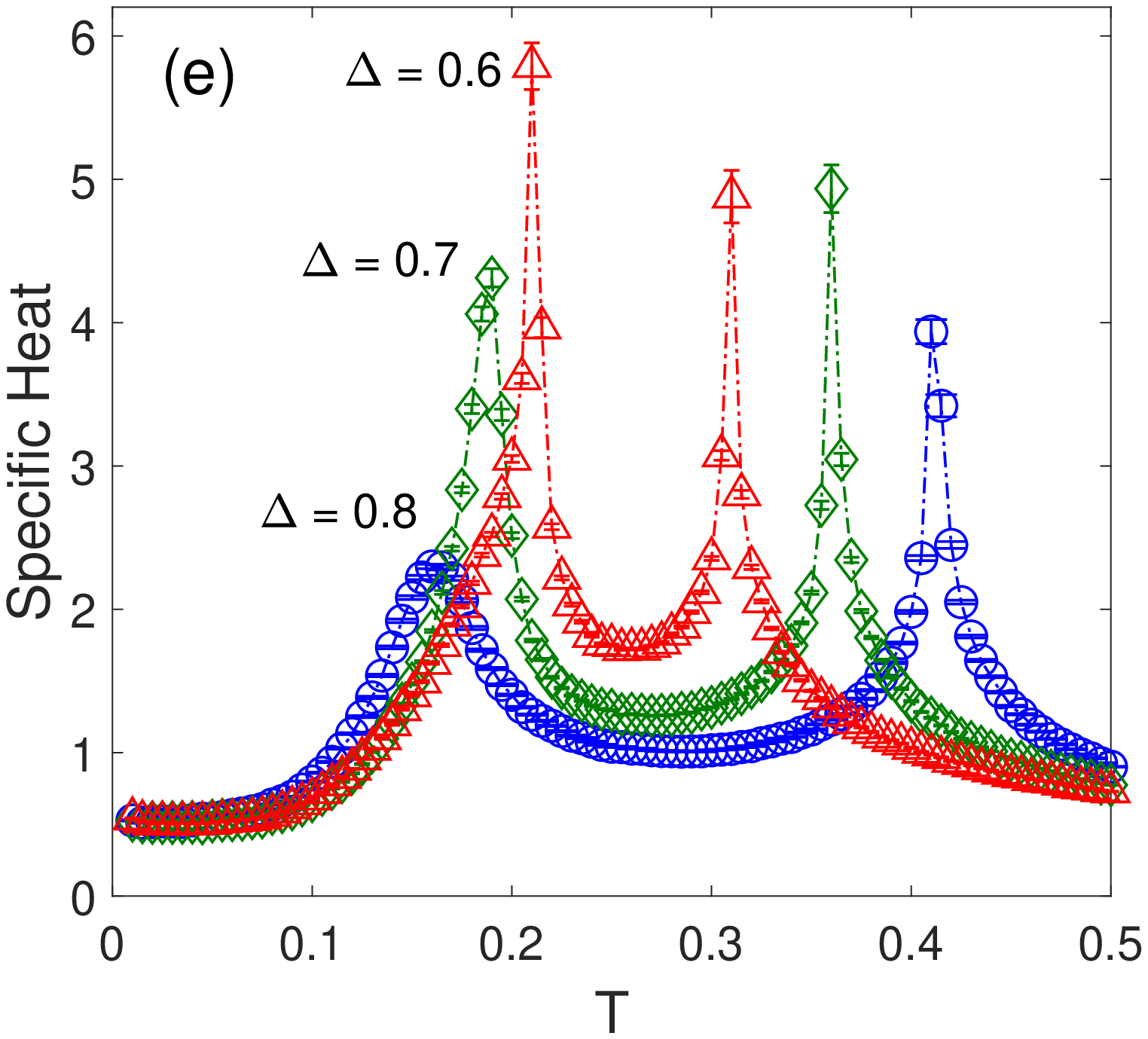}\label{fig:C_q7_x08}}
\subfigure{\includegraphics[scale=0.34,clip]{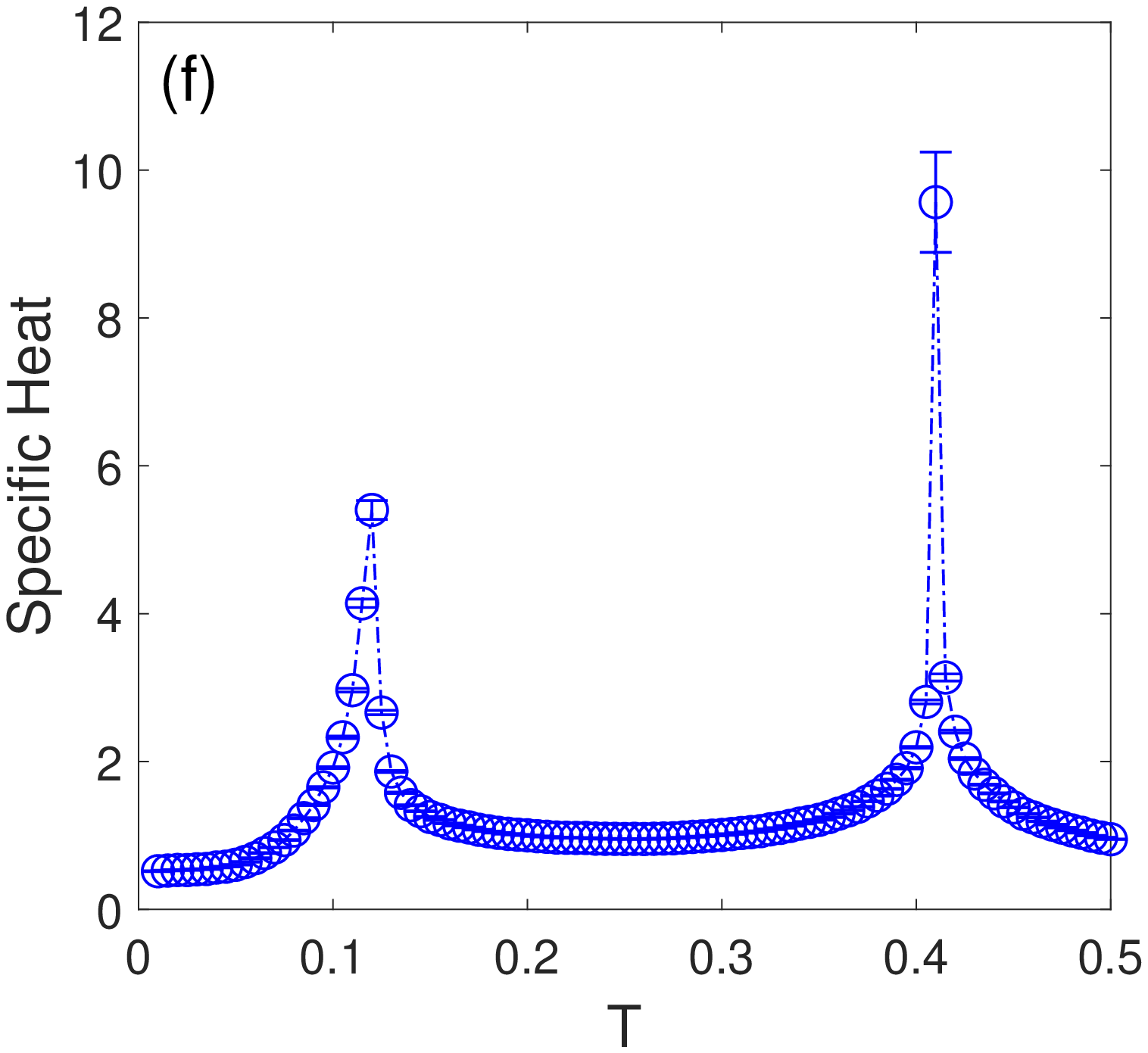}\label{fig:C_q10_x08}}
\caption{Temperature dependencies of the specific heat $c$ for $q=4$ (left column) , $q=7$ (middle column) and $q=10$ (right column), corresponding to $\Delta = 0.4$ (upper row) and $\Delta = 0.8$ (lower row). In (c) and (e) the effect of $\Delta$ is demonstrated by adding curves for two more values.}
\label{fig:spec_heat}
\end{figure}

Potential phase transitions between different phases can be detected from the peaks in the response functions. In Fig.~\ref{fig:spec_heat} we present temperature dependencies of the specific heat for $q=4,7$ and 10 in the regimes of the dominant ANq (for $\Delta=0.4$ in the upper panels) and AFM (for $\Delta=0.8$ in the lower panels) interactions. Focusing first on the case of $\Delta=0.4$, for $q=4$ (Fig.~\ref{fig:C_q4_x04}) one can observe two distinct sharp peaks, pointing to the presence of two phase transitions. However, for $q=7$ (Fig.~\ref{fig:C_q7_x04}) an additional rounder peak appears in the low-temperature region and with the increasing $q$ it shifts to lower temperatures (see Fig.~\ref{fig:C_q10_x04} for $q=10$). The third peak suggests the possibility of another phase transition and thus the existence of four different phases. On the other hand, the picture is rather different in the regime when the AFM coupling prevails. The lower row in Fig.~\ref{fig:spec_heat} shows that for $q=4$ the specific heat displays only one sharp peak at higher and a relatively broad shoulder at lower temperatures (Fig.~\ref{fig:C_q4_x08}). Nevertheless, with the increasing $q$, the broad shoulder evolves first to a round but distinct peak (see Fig.~\ref{fig:C_q7_x08} for $q=7$) and then to a sharp peak (see Fig.~\ref{fig:C_q10_x08} for $q=10$), typical for a phase transition. The low-temperature round peaks do not get sharper only with the increasing $q$ but also with the decreasing $\Delta$, as demonstrated in the inset of Fig.~\ref{fig:C_q10_x04} and in Fig.~\ref{fig:C_q7_x08}. The question whether the round peaks reflect some kind of phase transition will be addressed below. Nevertheless, if all the anomalies observed in the specific heat behavior signified different phase transitions then we would witness the change of the phase diagram topology with the increasing $q$ from the symmetric one with two phase transitions on each side of the interval $\Delta$ to the asymmetric one with three phase transitions for smaller and two phase transitions for larger $\Delta$. Below we will try to shed more light on the critical behavior associated with the presented anomalies in the specific heat and clarify the nature of the corresponding phases.

\begin{figure}[t!]
\centering
\includegraphics[scale=0.5,clip]{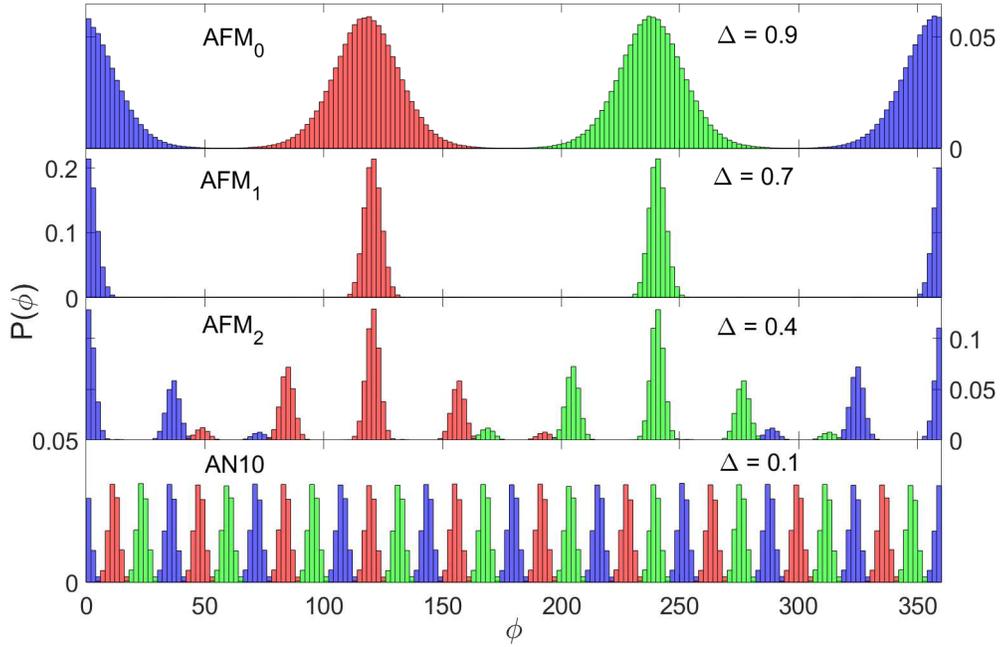}
\caption{Distributions of spin orientations in the observed phases for $q = 10$ at the fixed $T = 0.09$ and varying values of $\Delta$, obtained from a single configuration snapshot with $L = 1536$. Different colors (shades) represent different sublattices.}
\label{fig:Absolute_angles}
\end{figure}

In order to understand the character of the spin ordering in the possible different phases, separated by the specific heat anomalies, let us study spin distributions of microstates in the respective regions of the parameter space. Let us consider the case of $q=10$, for which the observed anomalies appear the most pronounced. To capture the lowest-temperature phase in the three-phase-transitions scenario at smaller $\Delta$ and study its disappearance/transition into its counterpart in the two-phase-transitions structure at larger $\Delta$, we fix the temperature to $T=0.09$ and vary the value of $\Delta$. In the lowest panel of Fig.~\ref{fig:Absolute_angles} we present the distributions of spin orientations (angles) for $\Delta=0.1$, i.e., in the limit of the strong ANq interaction. One can easily confirm that spins in each sublattice show $q=10$ possible orientations distributed with equal weights and the spin angle periodicity of $2\pi/10$, as one would expect in the ANq phase. The appropriate order parameter (for finite systems) in this phase is thus $m_{10}$. With the increasing influence of the AFM coupling the distribution undergoes a qualitative change, as demonstrated for $\Delta=0.4$ in the second panel from the bottom. Spins in each sublattice display only five preferential orientations with different weights, which are confined to the same half plane with the corresponding modes in different sublattices separated by the angle $2\pi/3$. Consequently, in each sublattice there is a net magnetization and the resulting AFM ordering between sublattices. 

\begin{figure}[t!]
\centering
\includegraphics[scale=0.5,clip]{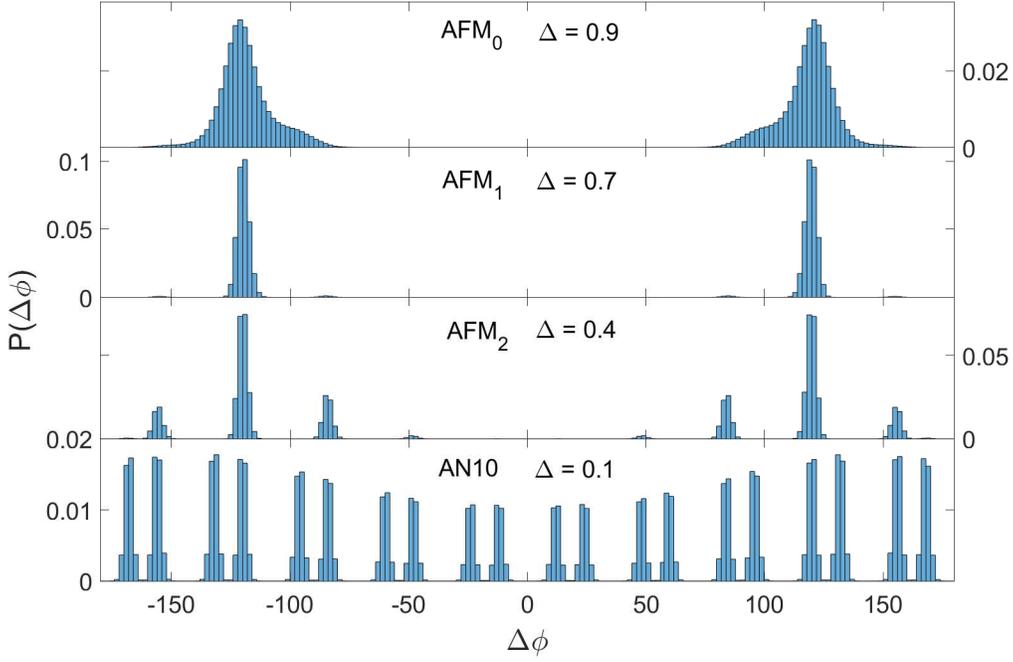}
\caption{Distributions of relative angles between neighboring spins in the observed phases for $q = 10$ at the fixed $T = 0.09$ and varying values of $\Delta$, obtained from a single configuration snapshot with $L = 1536$.}
\label{fig:Relative_angles}
\end{figure}

With further increase of the AFM coupling the possible spin orientations become more constrained with a single preferential direction in each sublattice, as shown for $\Delta=0.7$ in the second panel from the top. It is important to note that even though the resulting ordering is antiferromagnetic, it differs from the standard AFM phase in the absence of the higher-order coupling. In particular, owing to the persisting effect of the ANq coupling the widths of the sublattice spin distributions are constrained by the value $2\pi/q$. Considering the above arguments, the appropriate order parameter for the transition between these two peculiar AFM phases is $m_5$. Finally, in the strong limit of the AFM coupling, such as that for $\Delta=0.9$ presented in the top panel of Fig.~\ref{fig:Absolute_angles}, the ordering becomes usual AFM, for which the spin distribution widths are only controlled by the temperature and the appropriate order parameter is $m_1$. One can notice that in this phase the distributions become much wider than for $\Delta=0.7$, even though the temperature remains the same.

In Fig.~\ref{fig:Relative_angles} we present the distributions of relative spin angles, $\Delta \phi = \phi_i - \phi_j$, between neighboring spins for the same parameters as in Fig.~\ref{fig:Absolute_angles}, which provide more information about local spin arrangements. In particular, one can clearly observe that the influence of even a relatively small ANq coupling (for $\Delta=0.9$ in the upper panel) or AFM coupling (for $\Delta=0.1$ in the bottom panel) can distort the symmetric distributions around $\Delta \phi = \pm 2\pi/3$ in the former case and reweigh and shift the equally weighted distributions around $\Delta \phi = \pm k\pi/q$, $k=1,\hdots,q$ in the latter case. 

\begin{figure}[t!]
\centering
\subfigure{\includegraphics[scale=0.34,clip]{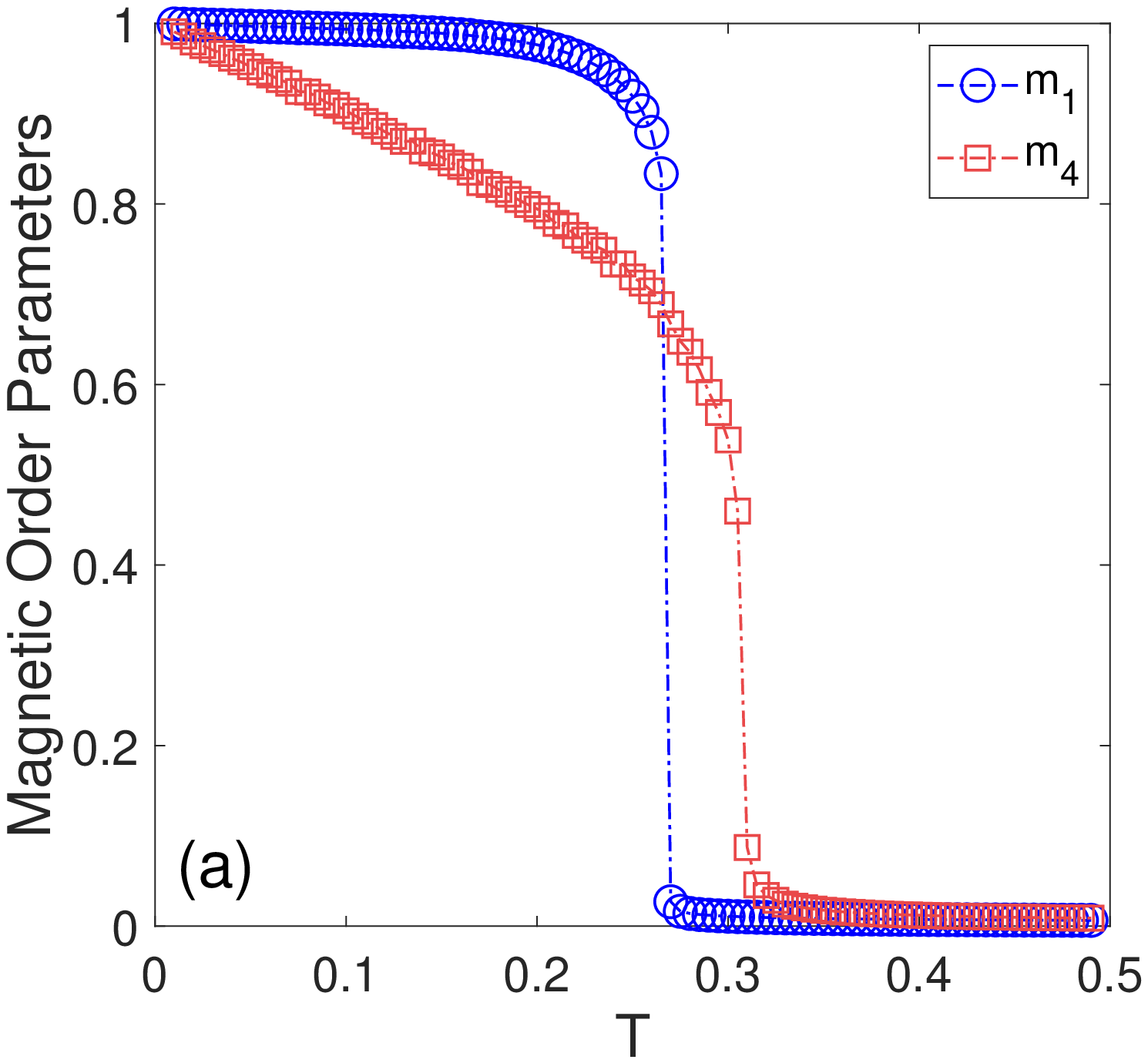}\label{fig:M_or_q4_x04}}
\subfigure{\includegraphics[scale=0.34,clip]{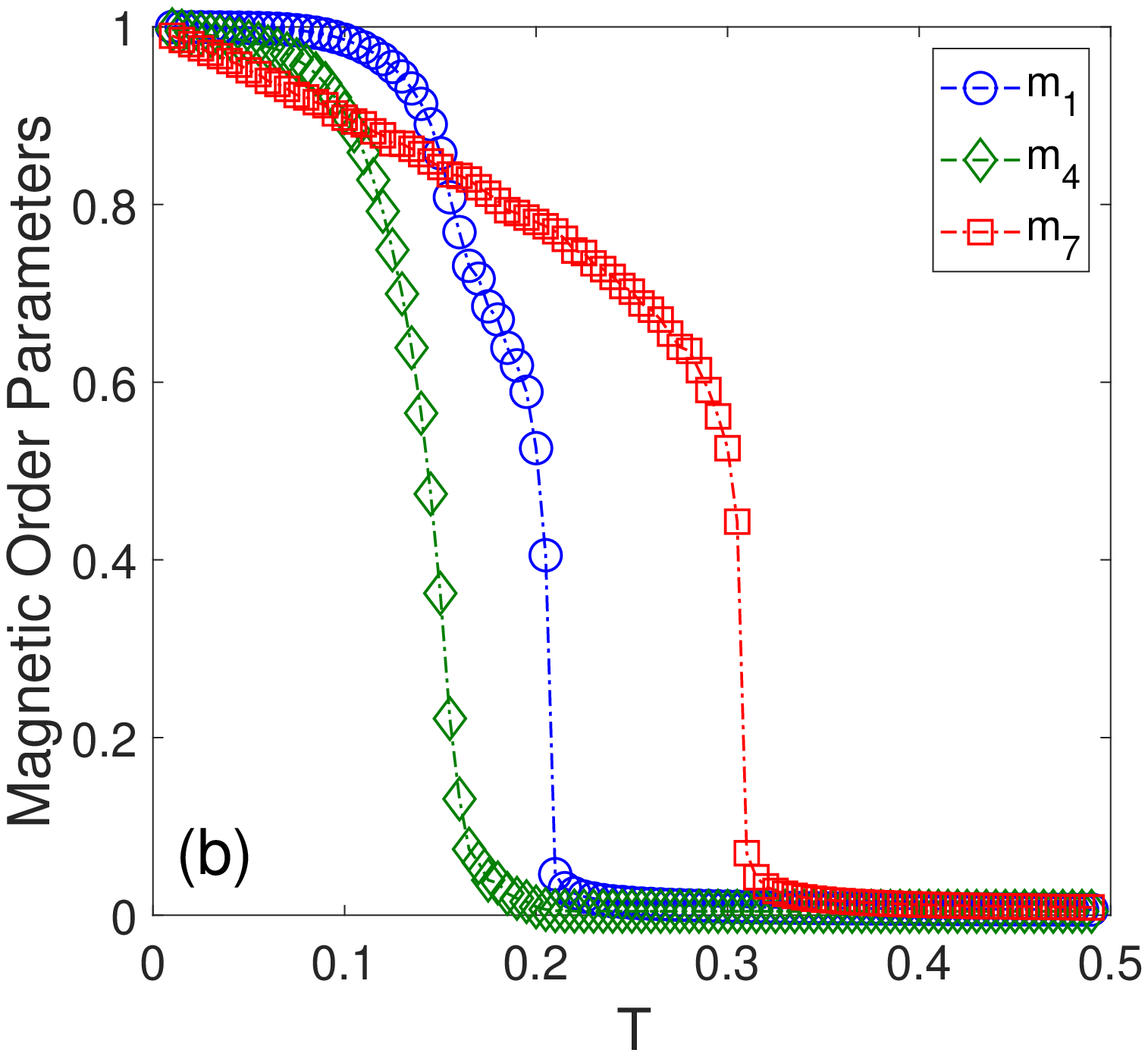}\label{fig:M_or_q7_x04}}
\subfigure{\includegraphics[scale=0.34,clip]{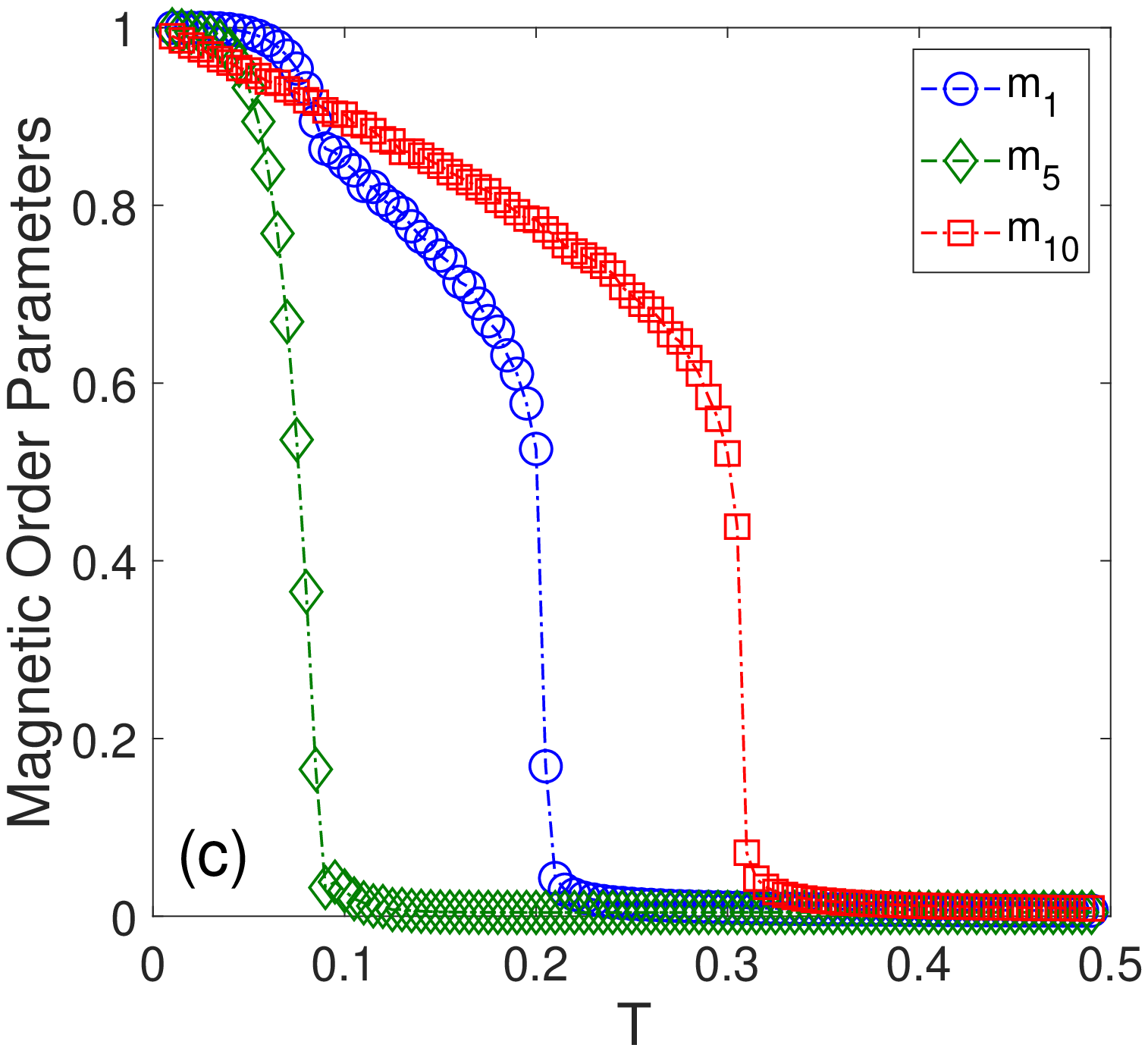}\label{fig:M_or_q10_x04}}\\
\subfigure{\includegraphics[scale=0.34,clip]{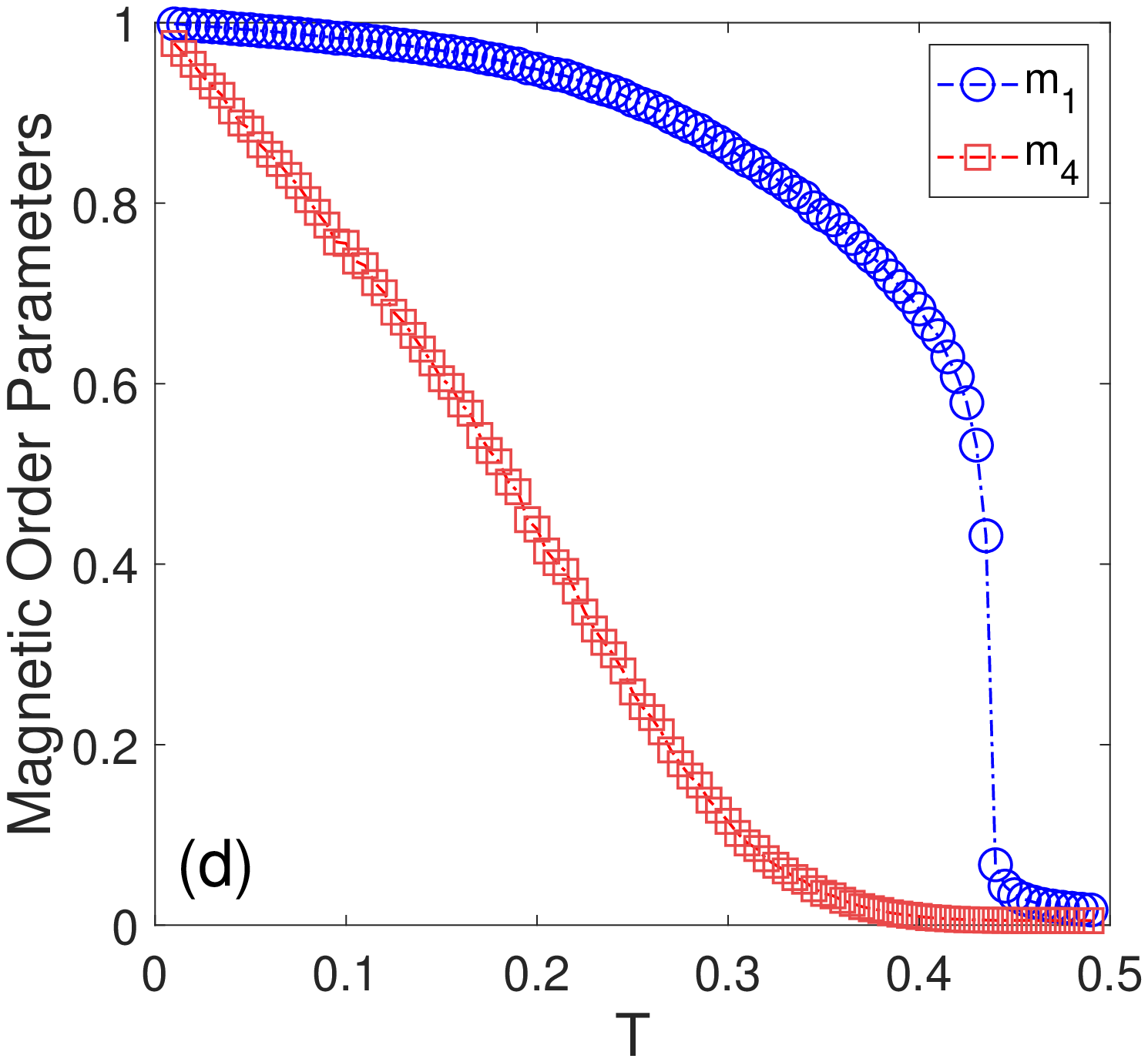}\label{fig:M_or_q4_x08}}
\subfigure{\includegraphics[scale=0.34,clip]{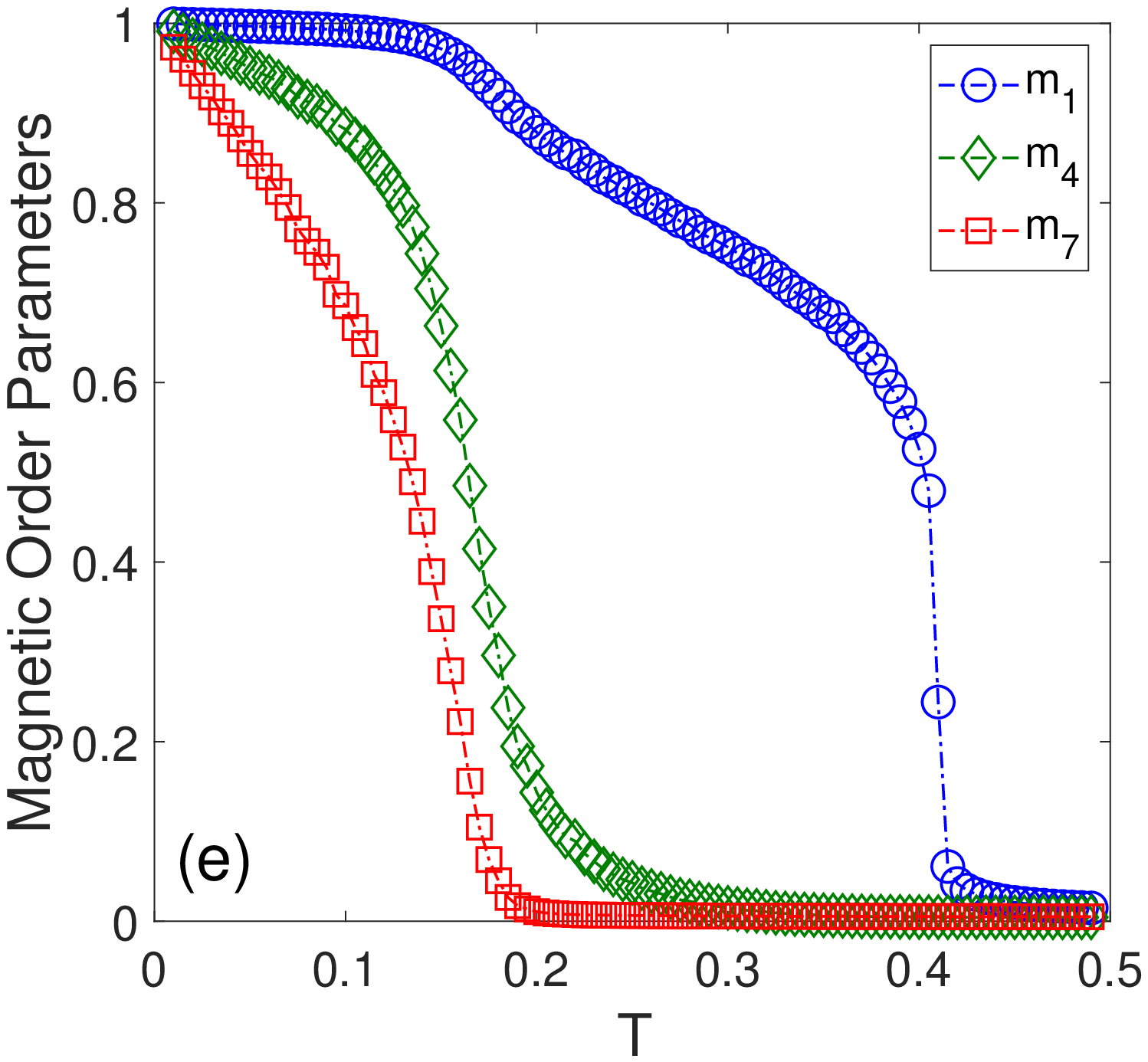}\label{fig:M_or_q7_x08}}
\subfigure{\includegraphics[scale=0.34,clip]{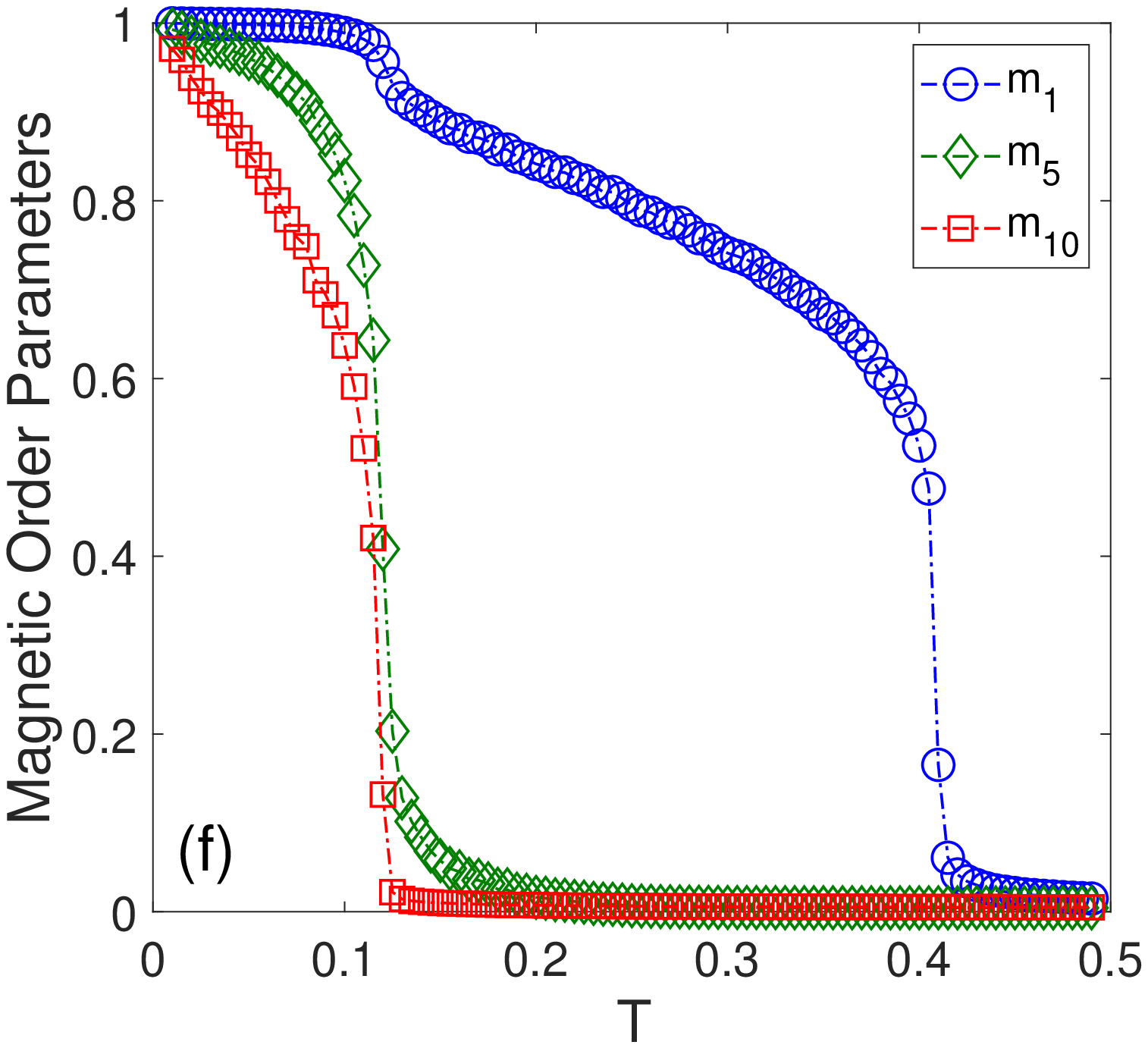}\label{fig:M_or_q10_x08}}
\caption{Temperature dependencies of the standard and generalized magnetic order parameters for $q=4$ (left column) , $q=7$ (middle column) and $q=10$ (right column), corresponding to $\Delta = 0.4$ (upper row) and $\Delta = 0.8$ (lower row).}
\label{fig:magn}
\end{figure}

\begin{figure}[t!]
\centering
\subfigure{\includegraphics[scale=0.34,clip]{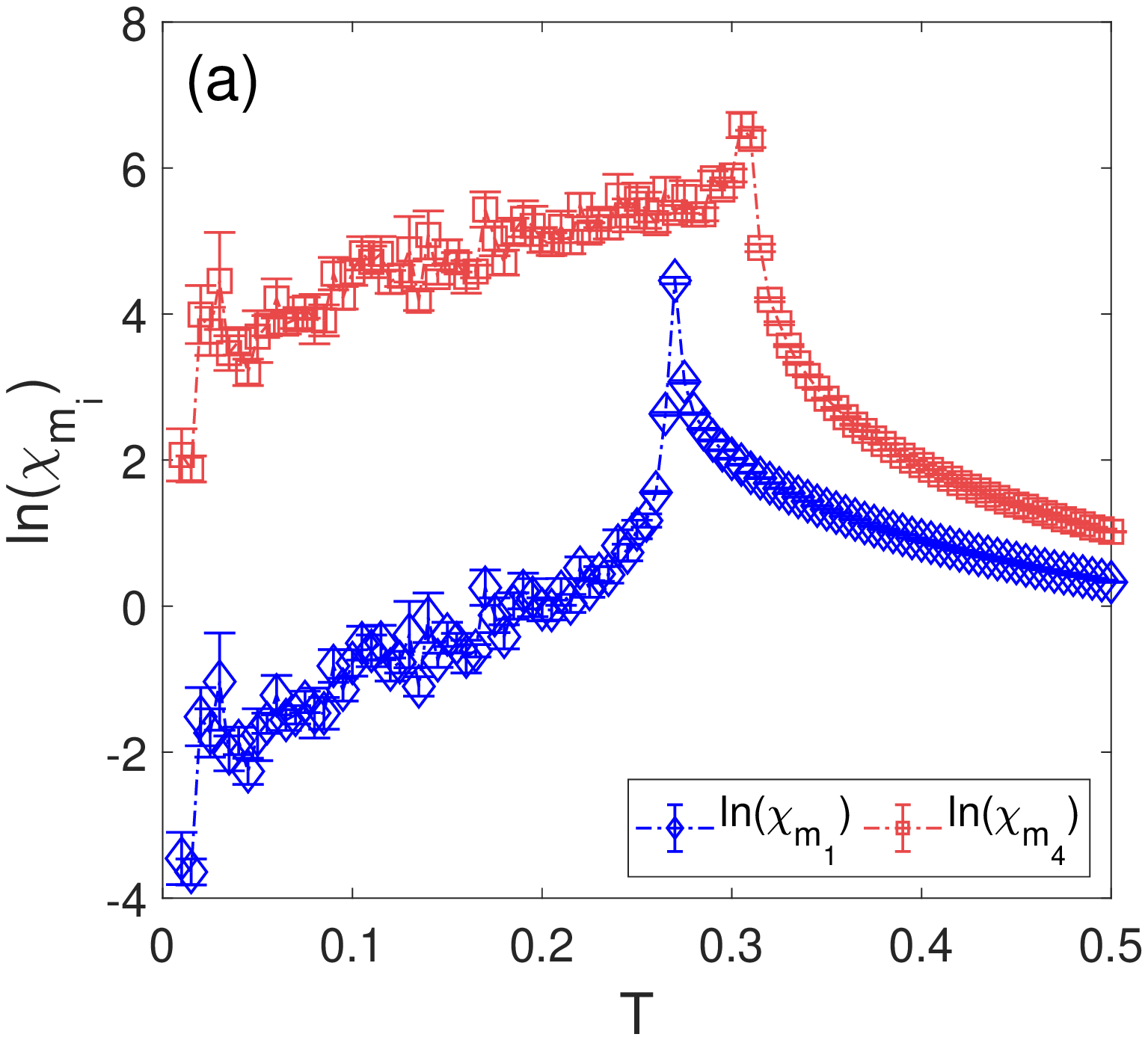}\label{fig:M_susc_q4_x04}}
\subfigure{\includegraphics[scale=0.34,clip]{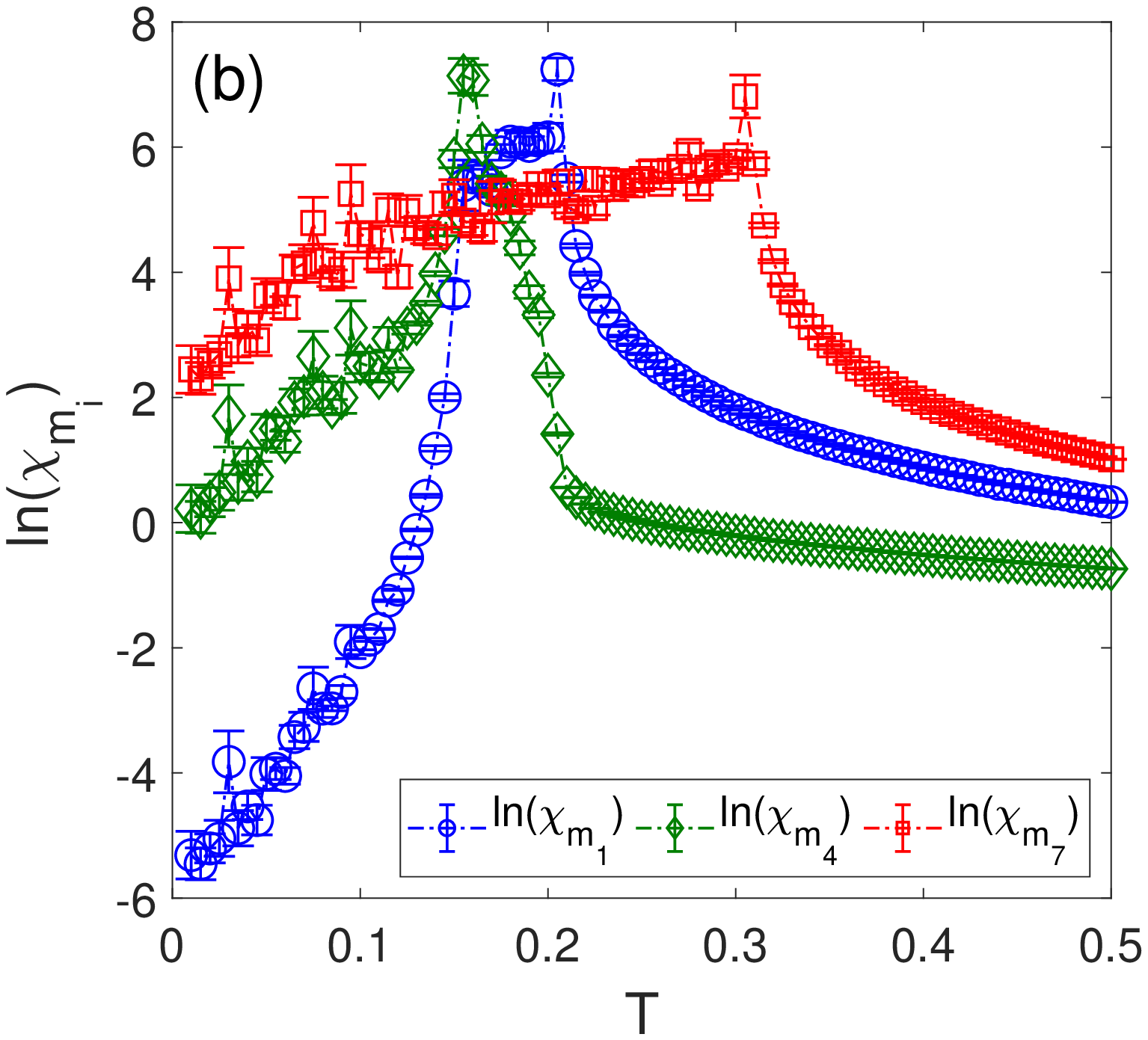}\label{fig:M_susc_q7_x04}}
\subfigure{\includegraphics[scale=0.34,clip]{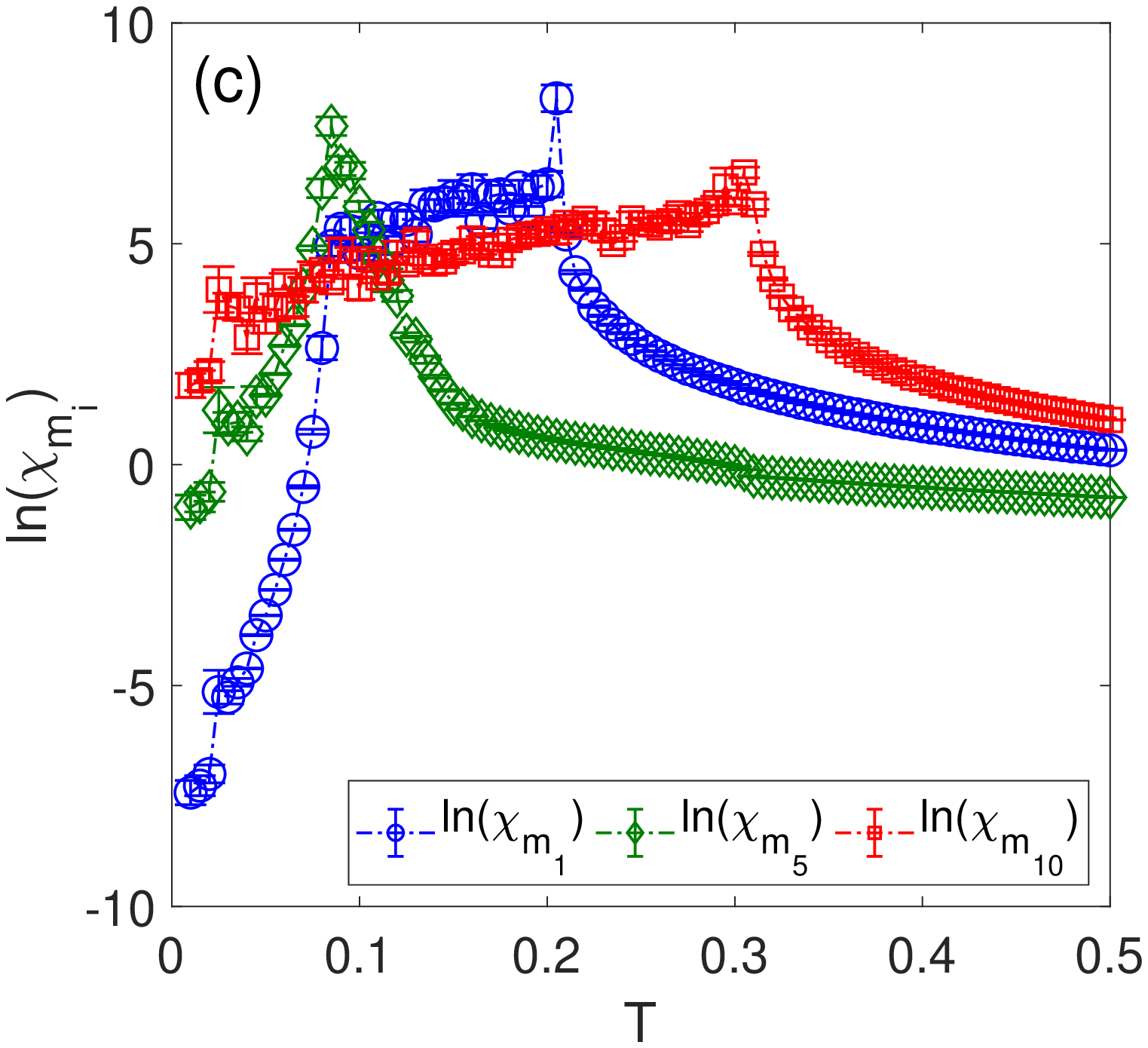}\label{fig:M_susc_q10_x04}}\\
\subfigure{\includegraphics[scale=0.34,clip]{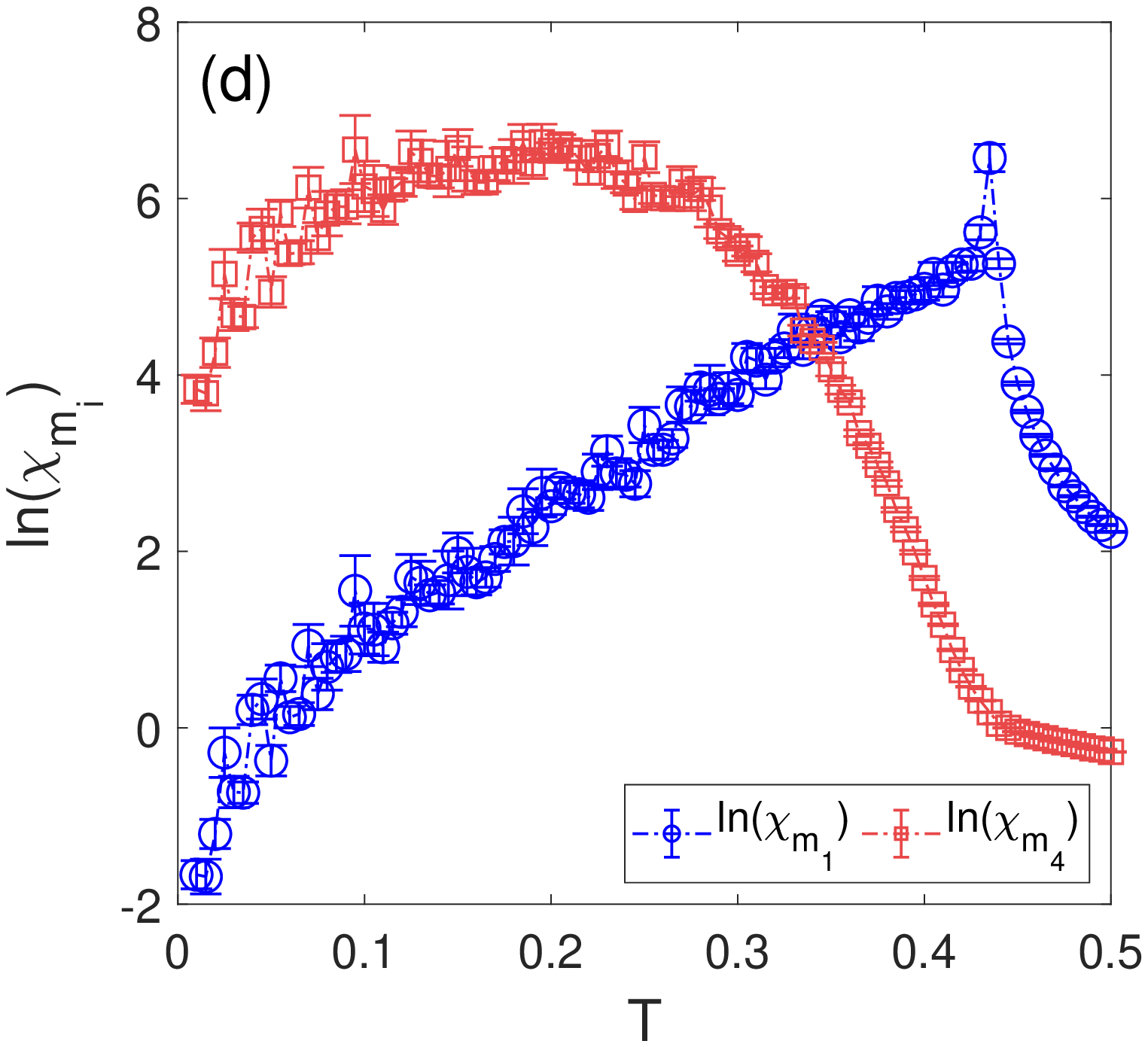}\label{fig:M_susc_q4_x08}}
\subfigure{\includegraphics[scale=0.34,clip]{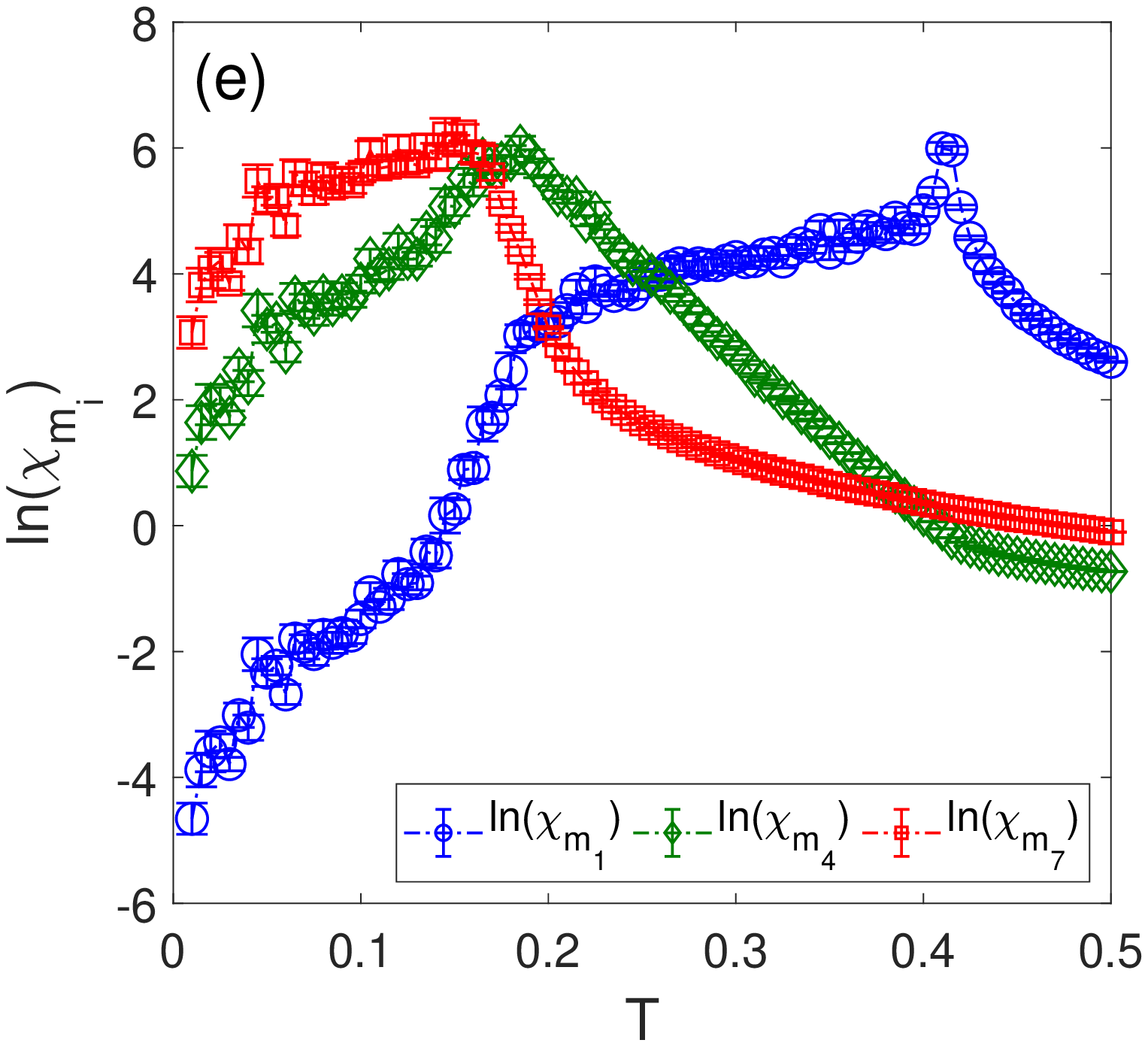}\label{fig:M_susc_q7_x08}}
\subfigure{\includegraphics[scale=0.34,clip]{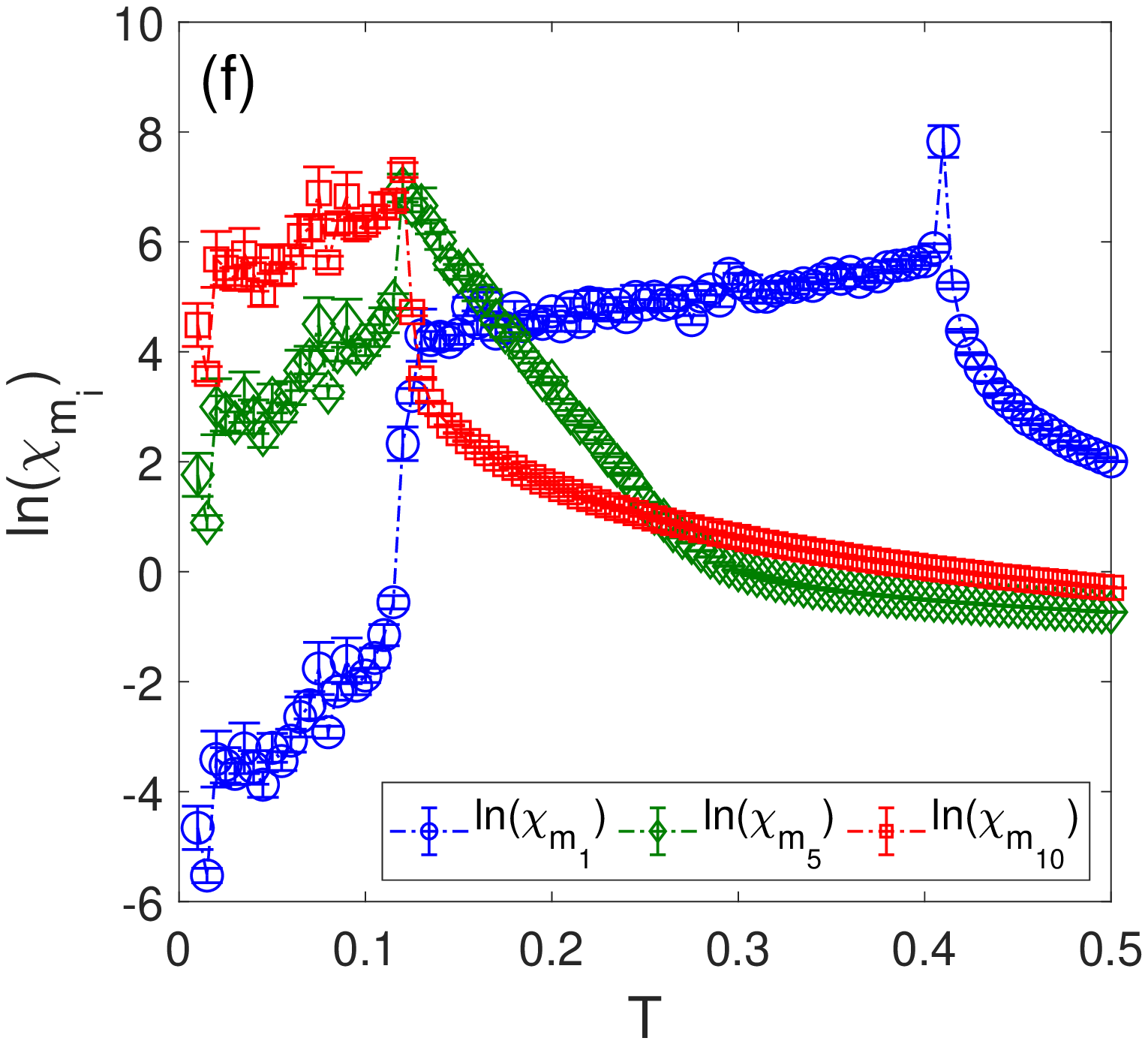}\label{fig:M_susc_q10_x08}}
\caption{Temperature dependencies of the standard and generalized magnetic susceptibilities for $q=4$ (left column) , $q=7$ (middle column) and $q=10$ (right column), corresponding to $\Delta = 0.4$ (upper row) and $\Delta = 0.8$ (lower row).}
\label{fig:magn_susc}
\end{figure}

Having characterized various phases and defined the appropriate parameters for magnetic and nematic ordering, let us examine their behavior, as well as the behavior of the corresponding response functions, in order to establish the respective phase boundaries. In Fig.~\ref{fig:magn} we present temperature variations of the relevant generalized magnetic order parameters and in Fig.~\ref{fig:magn_susc} the corresponding susceptibilities, for $q=4,7$ and 10, again in the regimes of the superior ANq ($\Delta=0.4$) and AFM ($\Delta=0.8$) interactions. 

For $\Delta=0.4$ (upper rows in Figs.~\ref{fig:magn} and~\ref{fig:magn_susc}), the respective order parameters, as well as the corresponding susceptibilities, indeed indicate two ($q=4$) and three ($q=7$ and 10) phase transitions, signaled by the anomalies in the specific heat above. From the order parameters it follows that for $q=4$ the system first displays the phase transition from the paramagnetic (P) to the AN4 phase, followed by another transition to the AFM phase at lower temperatures. The low-temperature phase thus features both the AN4 and AFM orderings. To distinguish it from the standard AFM\textsubscript{0} phase with no AN4 ordering, observed in the limit of large $\Delta$, we will refer to it as the AFM\textsubscript{1} phase. For $q=7$ and 10, another phase with a net AFM order emerges in between the AFM\textsubscript{1}, AFM\textsubscript{0} and ANq phases. This phase corresponds to the spin arrangement described in the third panel from the top in Figs.~\ref{fig:Absolute_angles} and~\ref{fig:Relative_angles} and will be referred to as AFM\textsubscript{2}.

For $\Delta=0.8$ (lower rows in Figs.~\ref{fig:magn} and~\ref{fig:magn_susc}), as expected, the order of the magnetic and nematic transitions is reversed: the former precedes the latter as the temperature is lowered. Again, in line with the prediction based on the specific heat behavior, only two possible phase transitions can be observed for all values of $q$. Furthermore, the broad shoulder appearing in the specific heat for $q=4$ translates in a very gentle decay of the order parameter $m_4$ and a broad peak of the associated susceptibility $\chi_{m_4}$. 

\begin{figure}[t!]
\centering
\subfigure{\includegraphics[scale=0.5,clip]{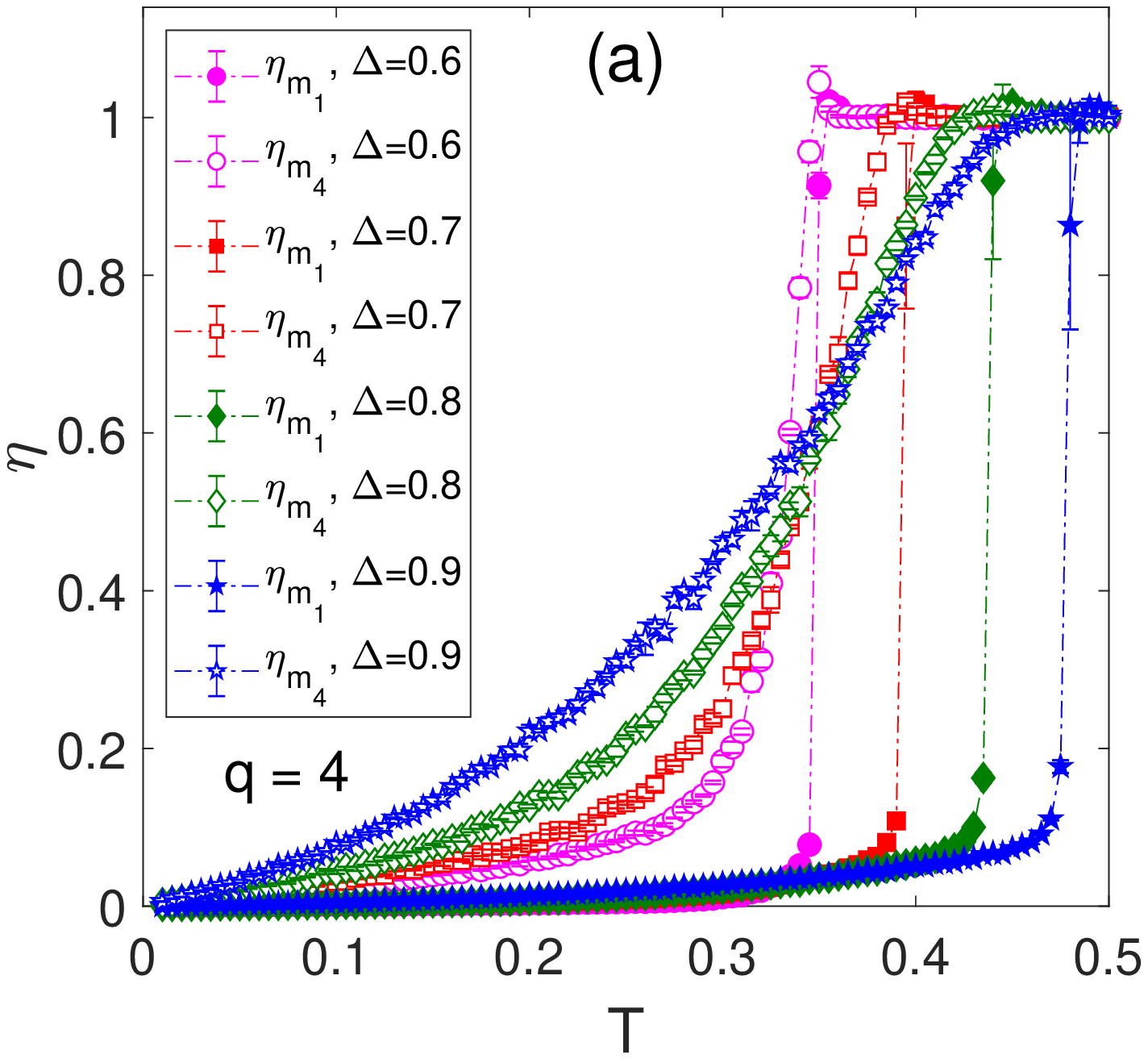}\label{fig:eta_m_q4}}
\subfigure{\includegraphics[scale=0.5,clip]{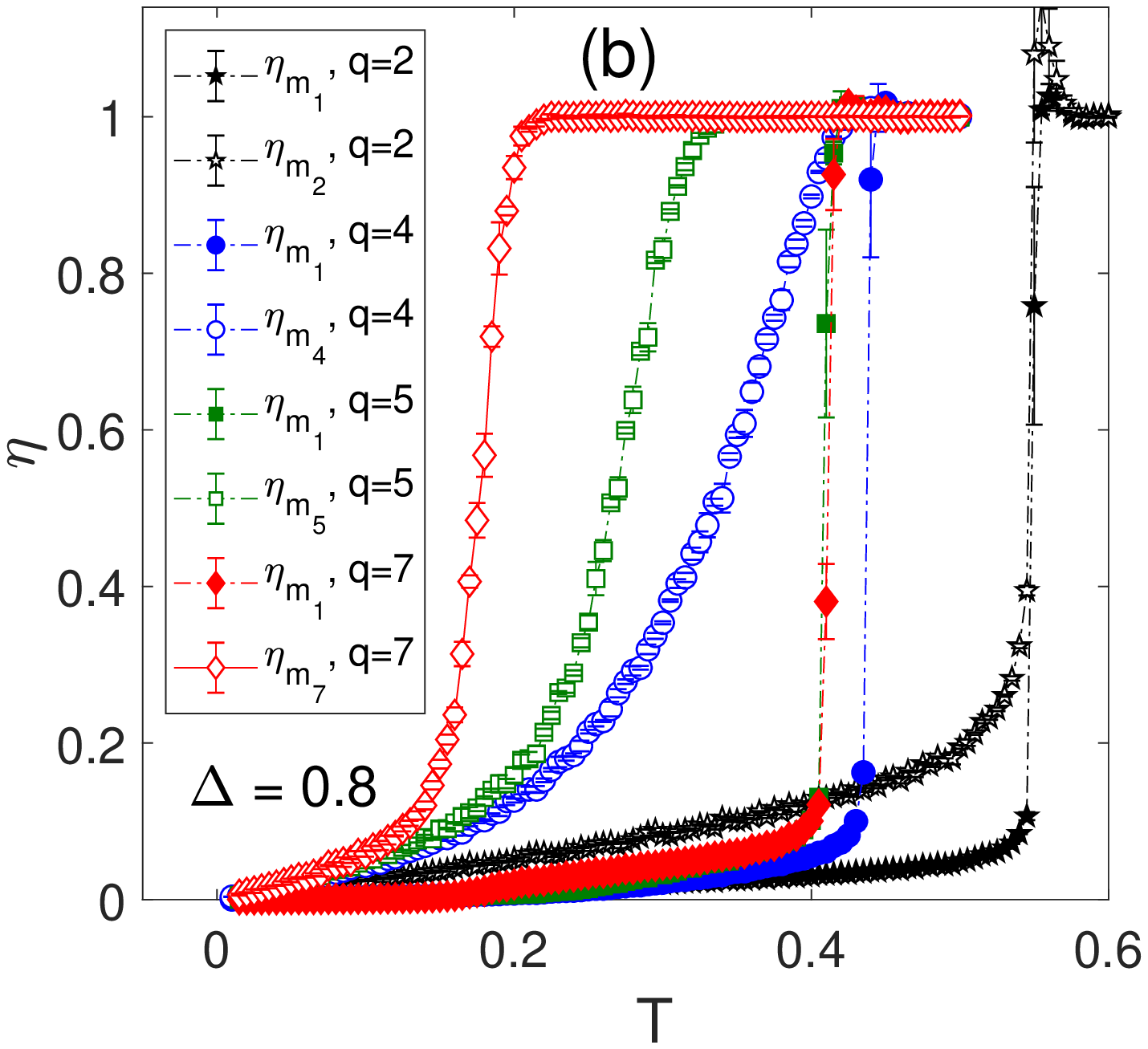}\label{fig:eta_m_delta08}}
\caption{Temperature dependencies of the critical exponents $\eta_{m_1}$ (full symbols) and $\eta_{m_q}$ (empty symbols), respectively, for (a) $q=4$ and different values of $\Delta$ and (b) $\Delta=0.8$ and different values of $q$.}
\label{fig:eta}
\end{figure}

Such a behavior is not typical for a phase transition and, therefore, to better explore it we further perform a FSS analysis, based on the scaling relation~(\ref{eq:fss_o_bkt}), and study the associated correlation functions. In order to distinguish the AFM\textsubscript{0} phase with solely magnetic algebraic correlations from the AFM\textsubscript{1} phase with both magnetic and generalized nematic correlations, we study decays of the pair-correlation functions $G_1$ and $G_q$. In Fig.~\ref{fig:eta_m_q4} we show for $q=4$ the temperature dependencies of the corresponding critical exponents $\eta_{m_1}$ (full symbols) and $\eta_{m_4}$ (empty symbols), respectively, for different values of $\Delta$. One can observe a sharp increase of $\eta_{m_1}$ at the AFM\textsubscript{0}-P phase transition but only a rather gentle increase of $\eta_{m_4}$ spread over a wide temperature interval. Nevertheless, at least for larger values of $\Delta$, it is apparent that the value of 1, corresponding to the exponential decay of the correlation function in the paramagnetic phase, is reached by $\eta_{m_4}$ at temperatures lower than those corresponding to $\eta_{m_1}$. This finding tells us that the generalized nematic (AN4) correlations disappear before the magnetic (AFM) ones and, thus, it suggests the existence of separate AFM\textsubscript{1} and AFM\textsubscript{0} phases. However, the behavior of $\eta_{m_4}$ is not typical for a standard phase transition and, thus, it might signal a crossover instead of a phase transition. 

\begin{figure}[t!]
\centering
\subfigure{\includegraphics[scale=0.34,clip]{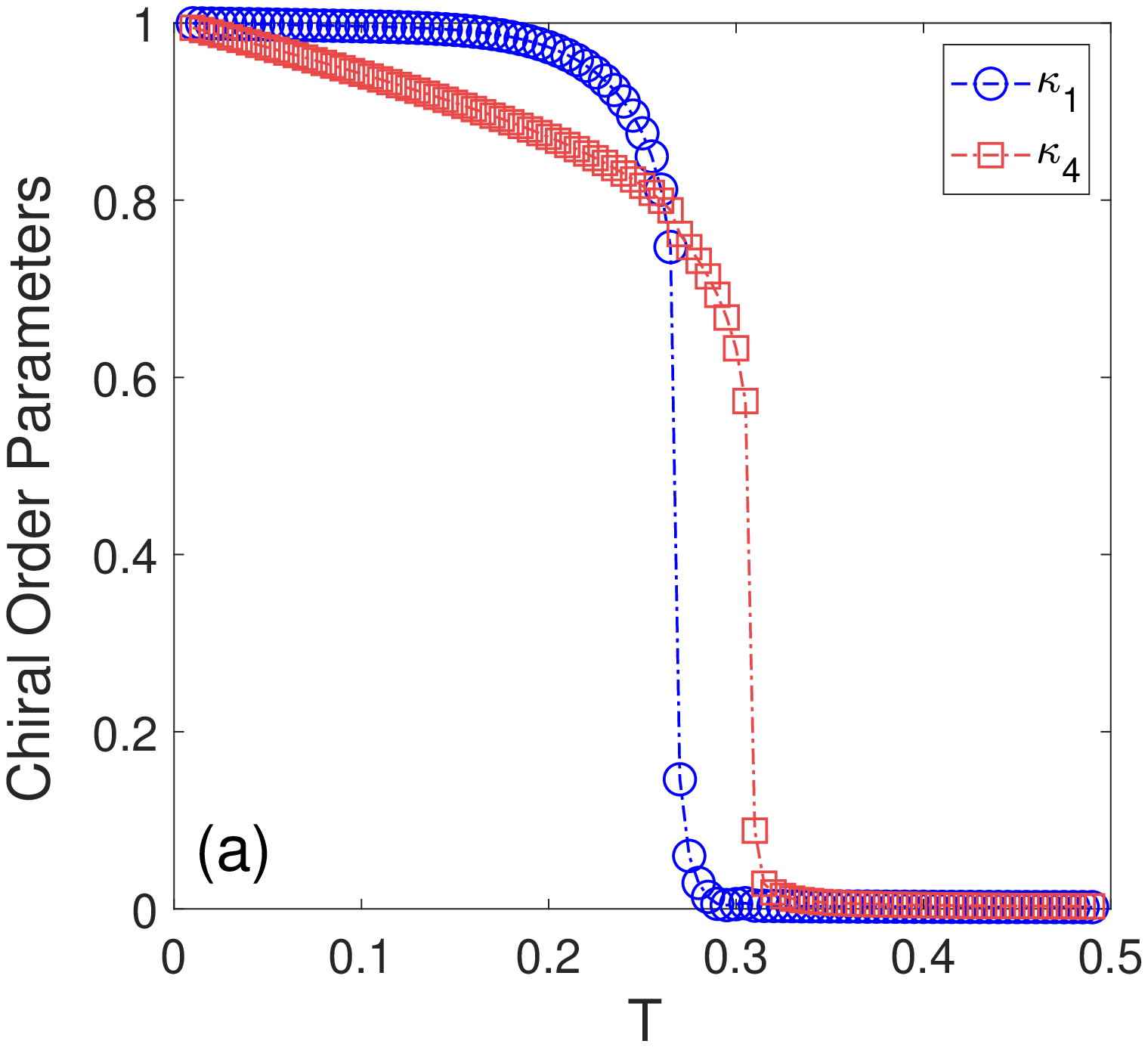}\label{fig:Chi_or_q4_x04}}
\subfigure{\includegraphics[scale=0.34,clip]{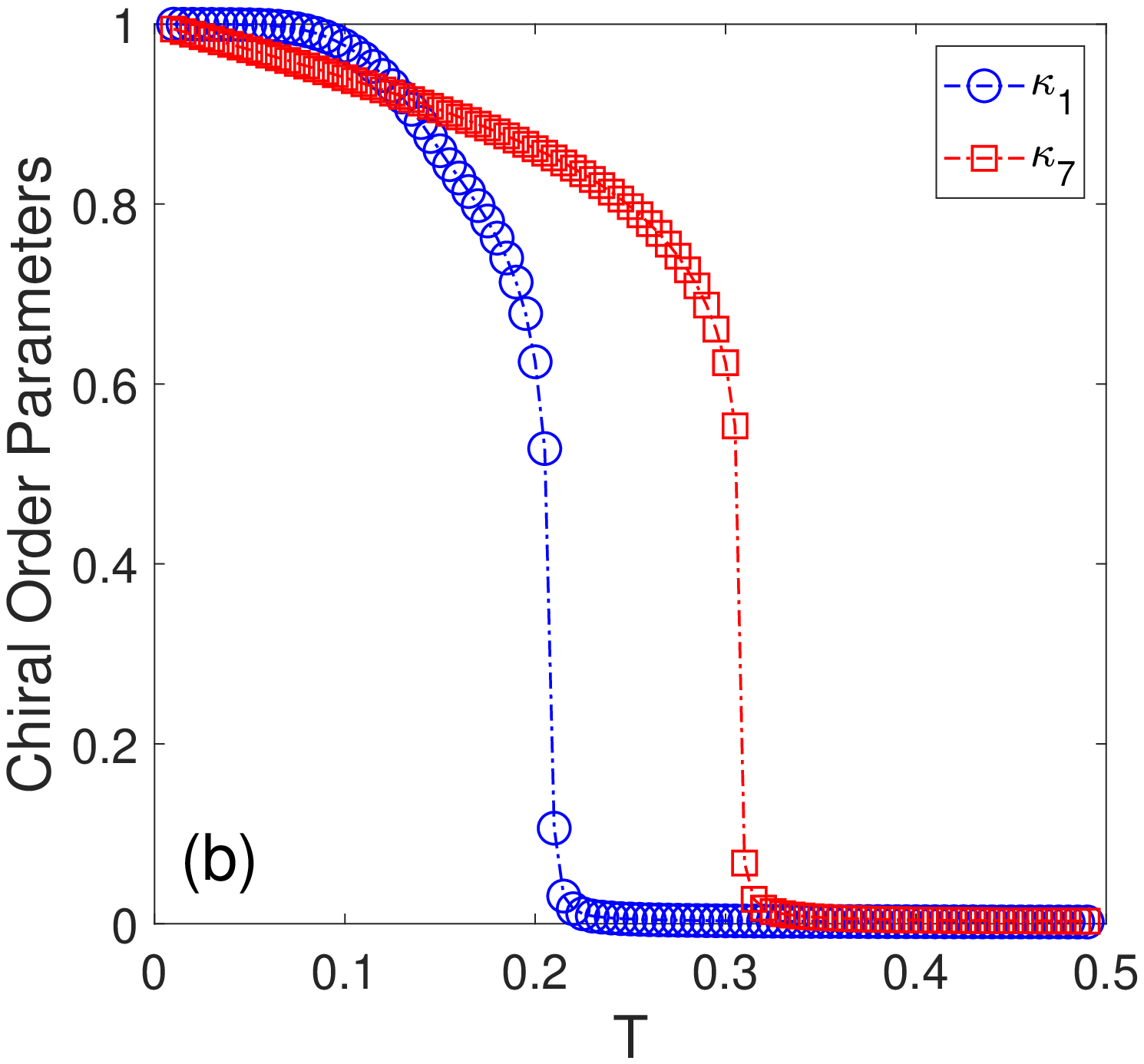}\label{fig:Chi_or_q7_x04}}
\subfigure{\includegraphics[scale=0.34,clip]{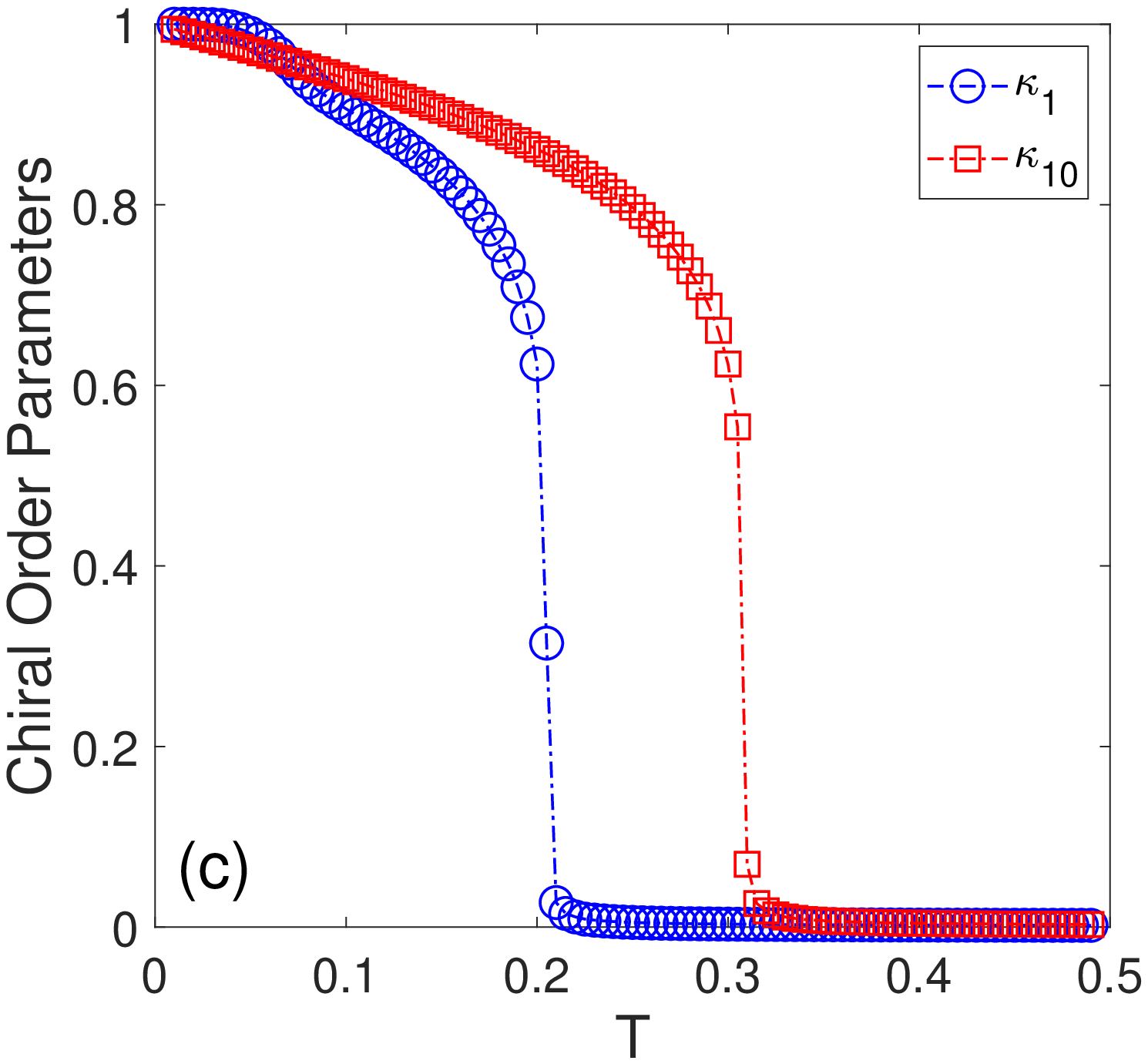}\label{fig:Chi_or_q10_x04}}\\
\subfigure{\includegraphics[scale=0.34,clip]{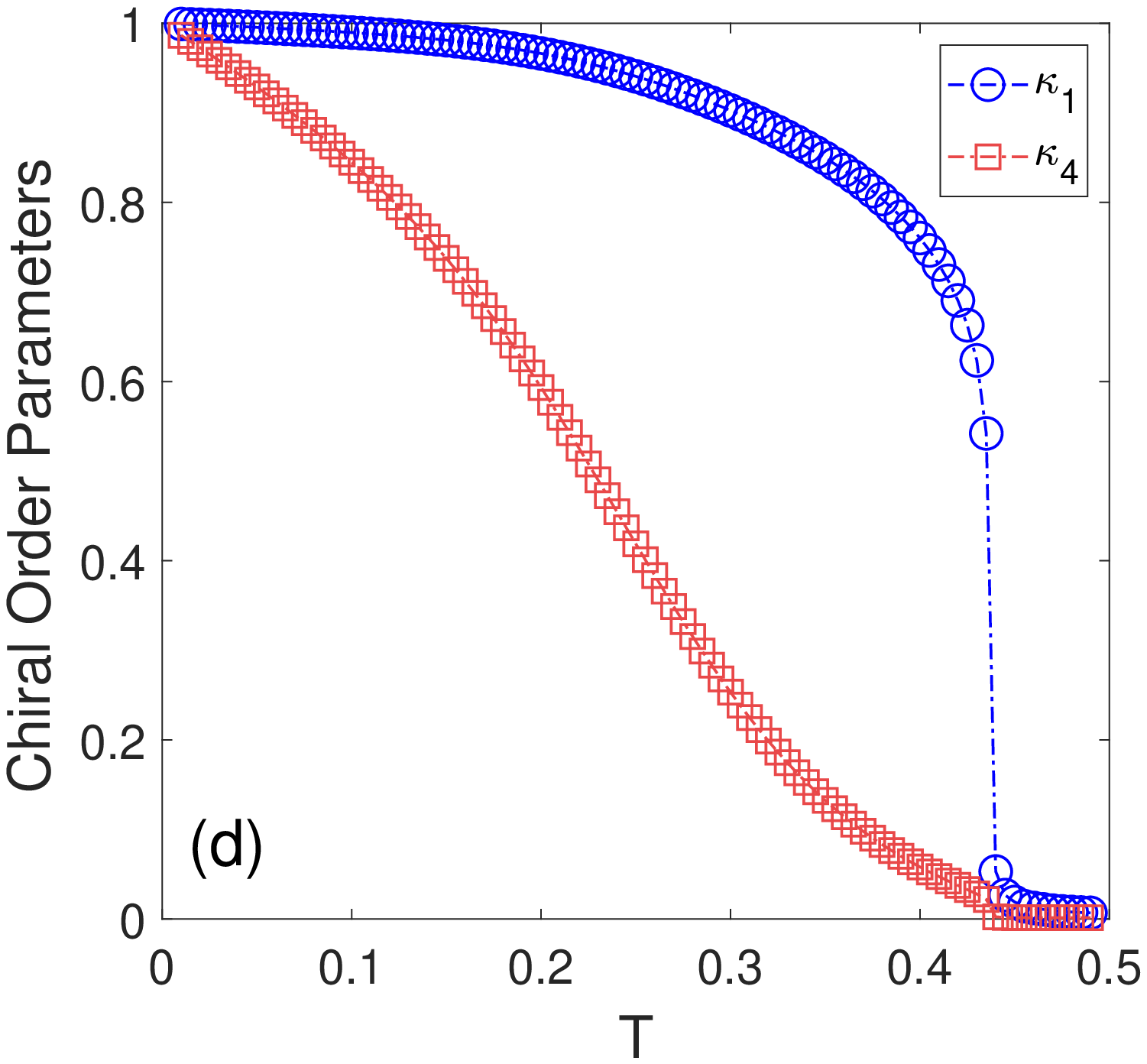}\label{fig:Chi_or_q4_x08}}
\subfigure{\includegraphics[scale=0.34,clip]{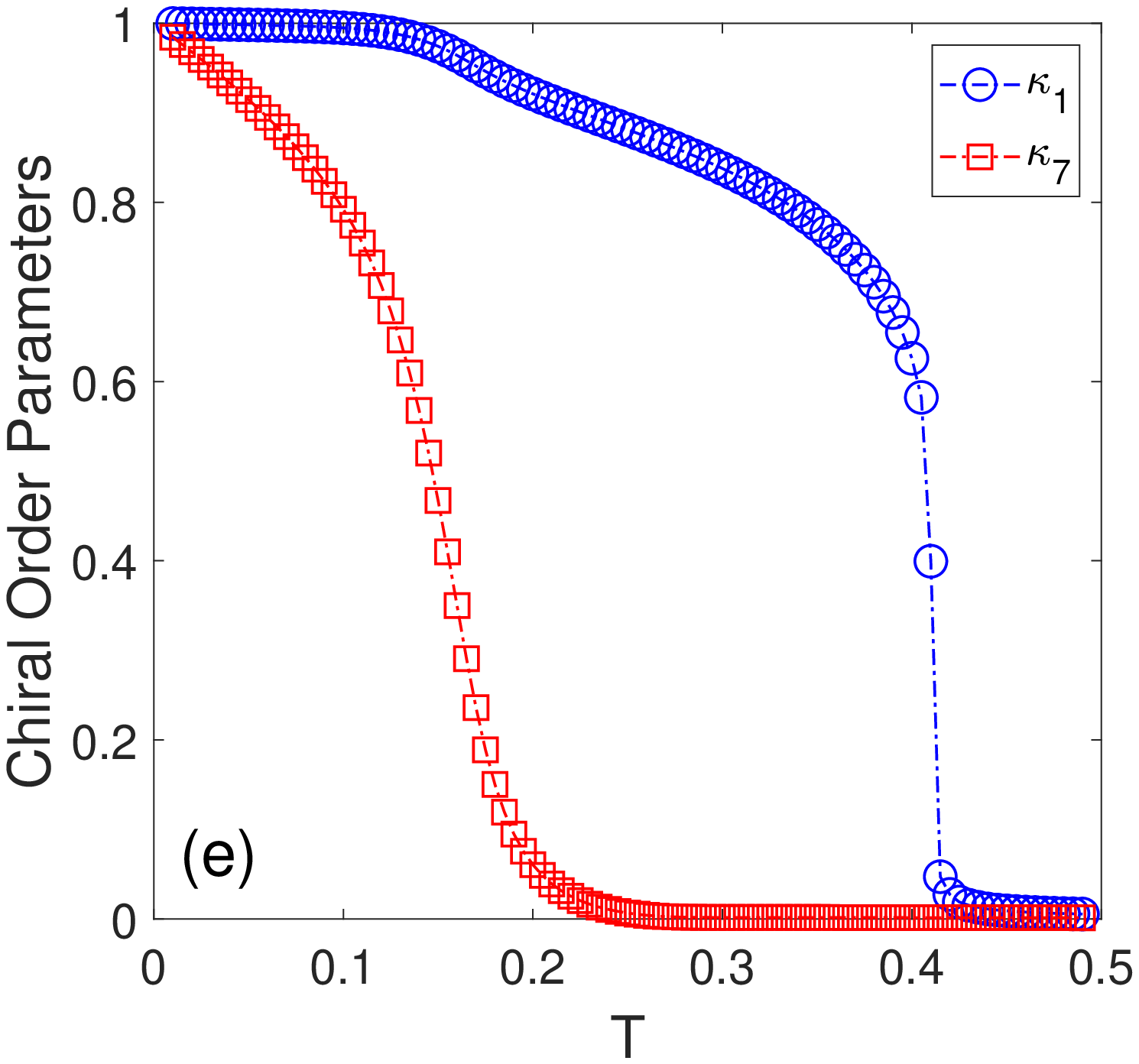}\label{fig:Chi_or_q7_x08}}
\subfigure{\includegraphics[scale=0.34,clip]{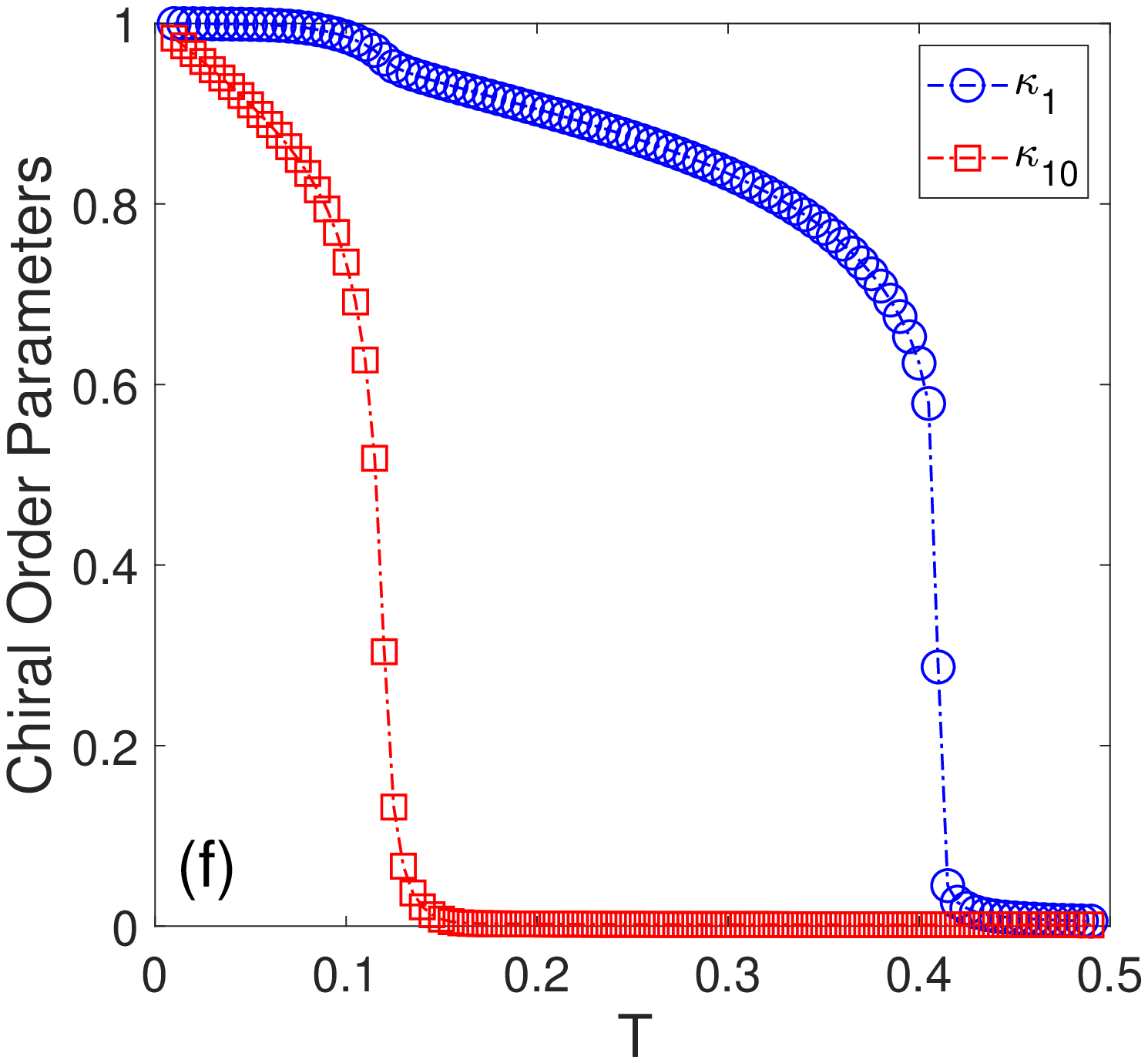}\label{fig:Chi_or_q10_x08}}
\caption{Temperature dependencies of the standard and generalized chiral order parameters for $q=4$ (left column) , $q=7$ (middle column) and $q=10$ (right column), corresponding to $\Delta = 0.4$ (upper row) and $\Delta = 0.8$ (lower row).}
\label{fig:chir}
\end{figure}

In Fig.~\ref{fig:eta_m_delta08} we demonstrate the separation of the AFM\textsubscript{0} and AFM\textsubscript{1} phases with the increasing $q \geq 2$ (nondivisible by 3) for a fixed $\Delta=0.8$. For $q=2$ the transition to the paramagnetic phase clearly occurs at the same temperature and thus the two phases do not separate. The separation becomes apparent for $q=4$ but the exponent $\eta_{m_4}$ crosses to the value of one gradually over an extended interval of temperatures. With further increase of $q$ the separation distance increases and the slope of $\eta_{m_q}$ becomes sharper. The shape of $\eta_{m_7}$ is already much closer to the standard phase transition behavior, nevertheless, the specific heat curves presented in Fig.~\ref{fig:spec_heat} would rather suggest that such a behavior with a sharp peak only occurs at still higher $q$. Our analysis confirms that for $\Delta=0.8$ it happens starting from $q=8$ (not shown). Nevertheless, as already indicated by the evolving shape of the specific heat peaks, the behavior of $\eta_{m_q}$ also depends on the value of $\Delta$. Consequently, for sufficiently low $\Delta \approx 0.6$ it is possible to obtain the standard critical behavior at the AFM\textsubscript{0}-AFM\textsubscript{1} phase transition with the sharp specific heat peak and the sharp increase of $\eta_{m_q}$ for the nematic parameter as low as $q=5$. For $q=4$ we were able to confirm by the correlation analysis the possibility of AFM\textsubscript{0} and AFM\textsubscript{1} phase separation but it does not seem to occur via a standard phase transition for any value of $\Delta$.

\begin{figure}[t!]
\centering
\subfigure{\includegraphics[scale=0.34,clip]{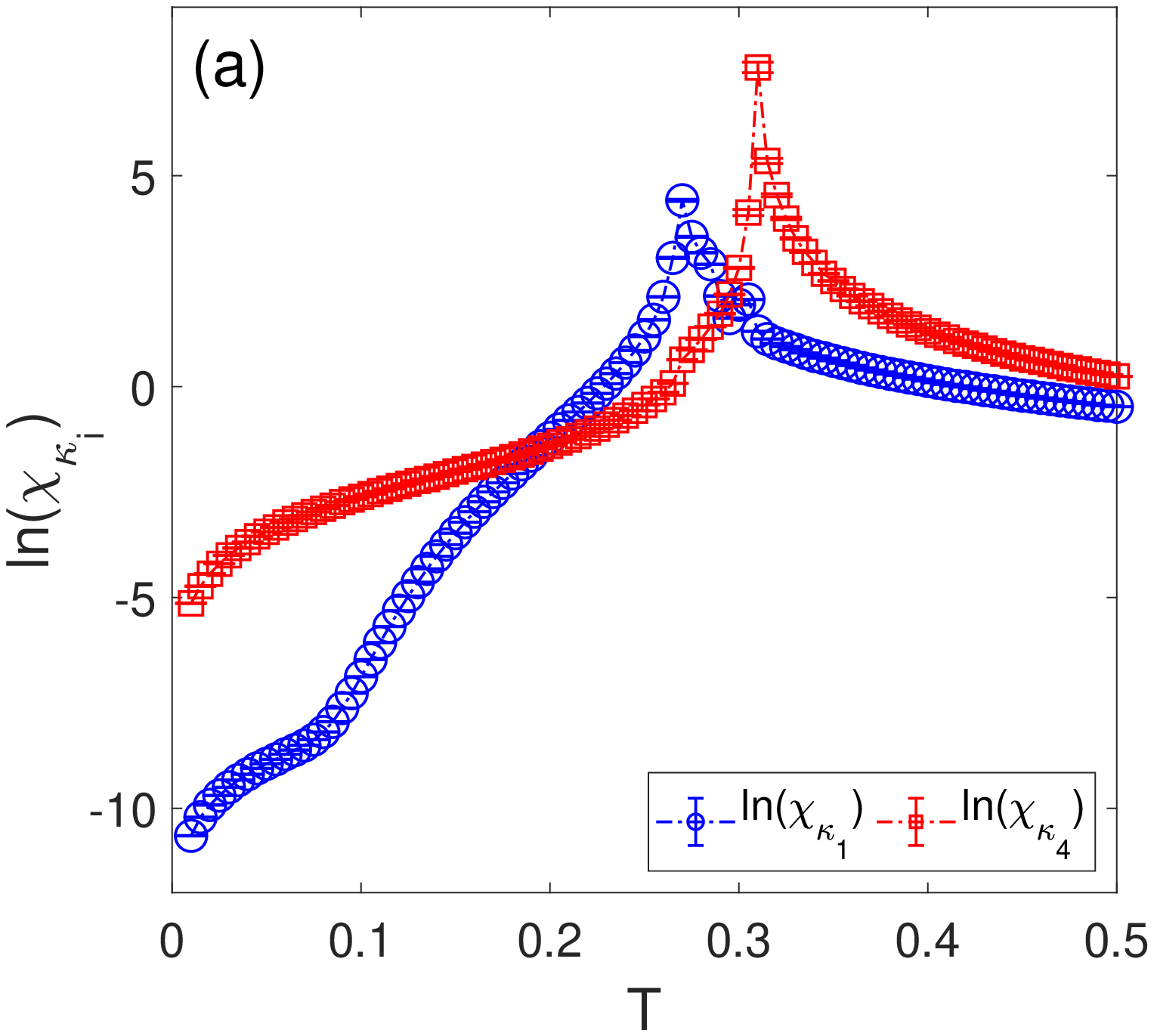}\label{fig:Chi_susc_q4_x04}}
\subfigure{\includegraphics[scale=0.34,clip]{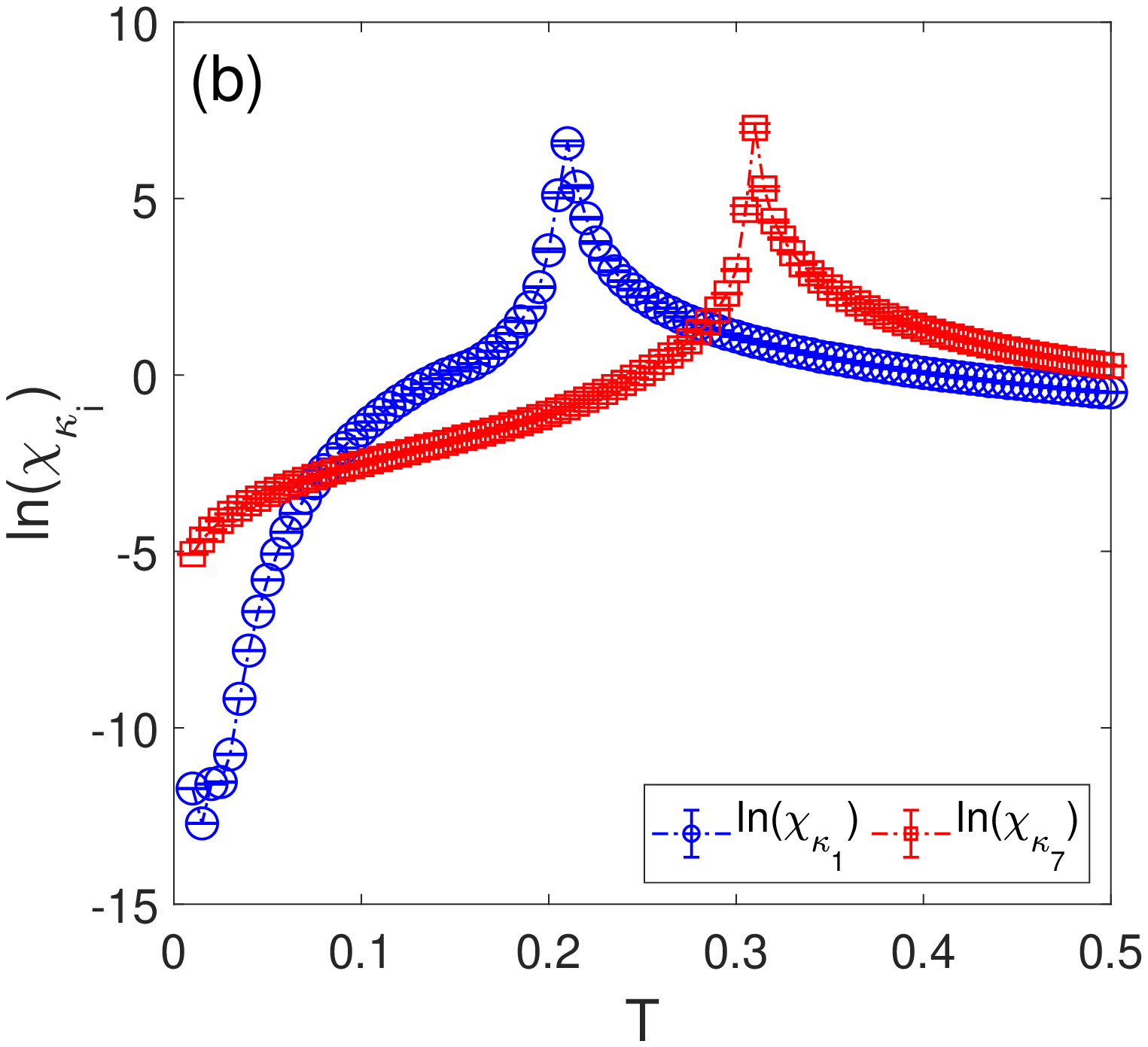}\label{fig:Chi_susc_q7_x04}}
\subfigure{\includegraphics[scale=0.34,clip]{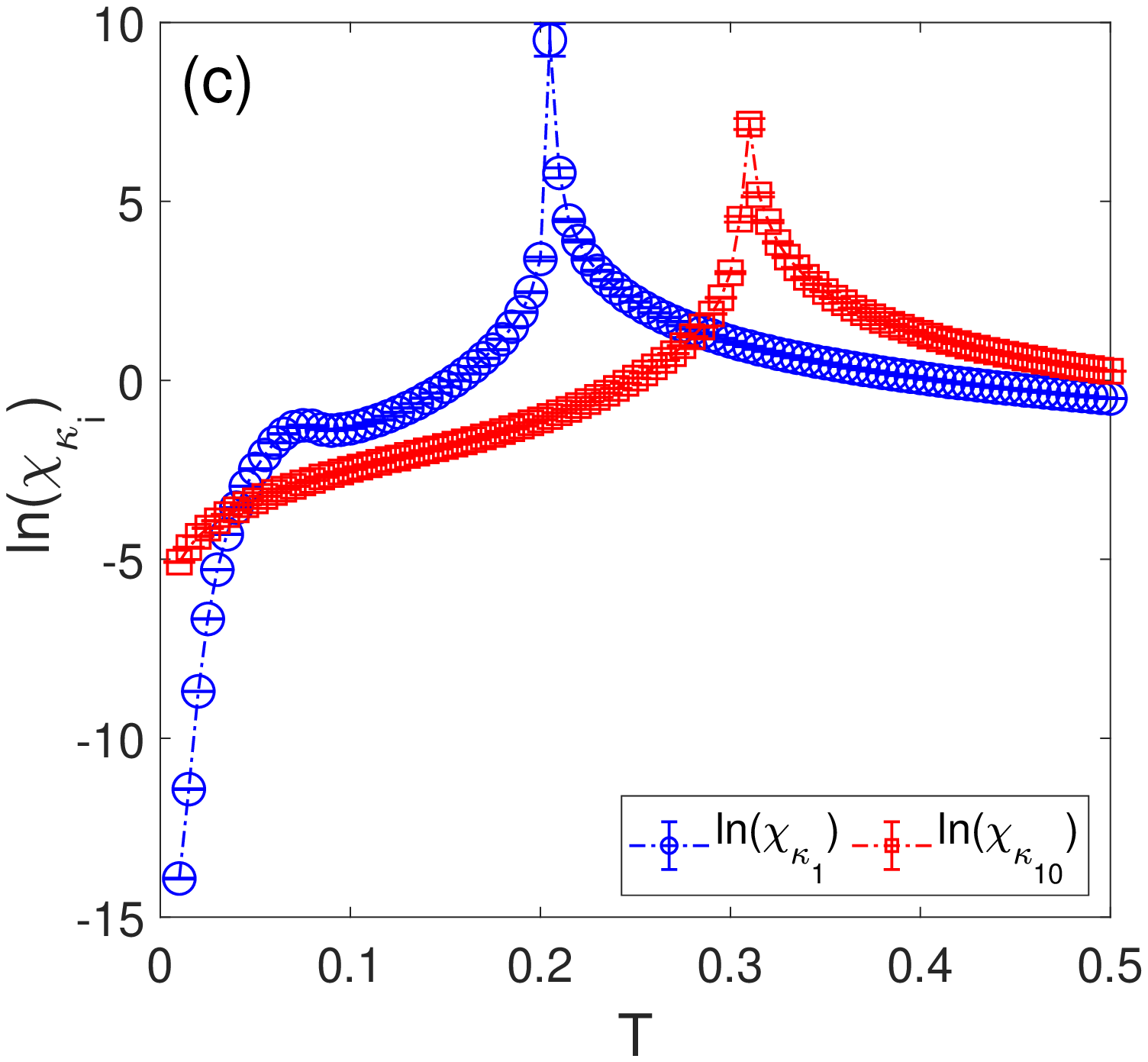}\label{fig:Chi_susc_q10_x04}}\\
\subfigure{\includegraphics[scale=0.34,clip]{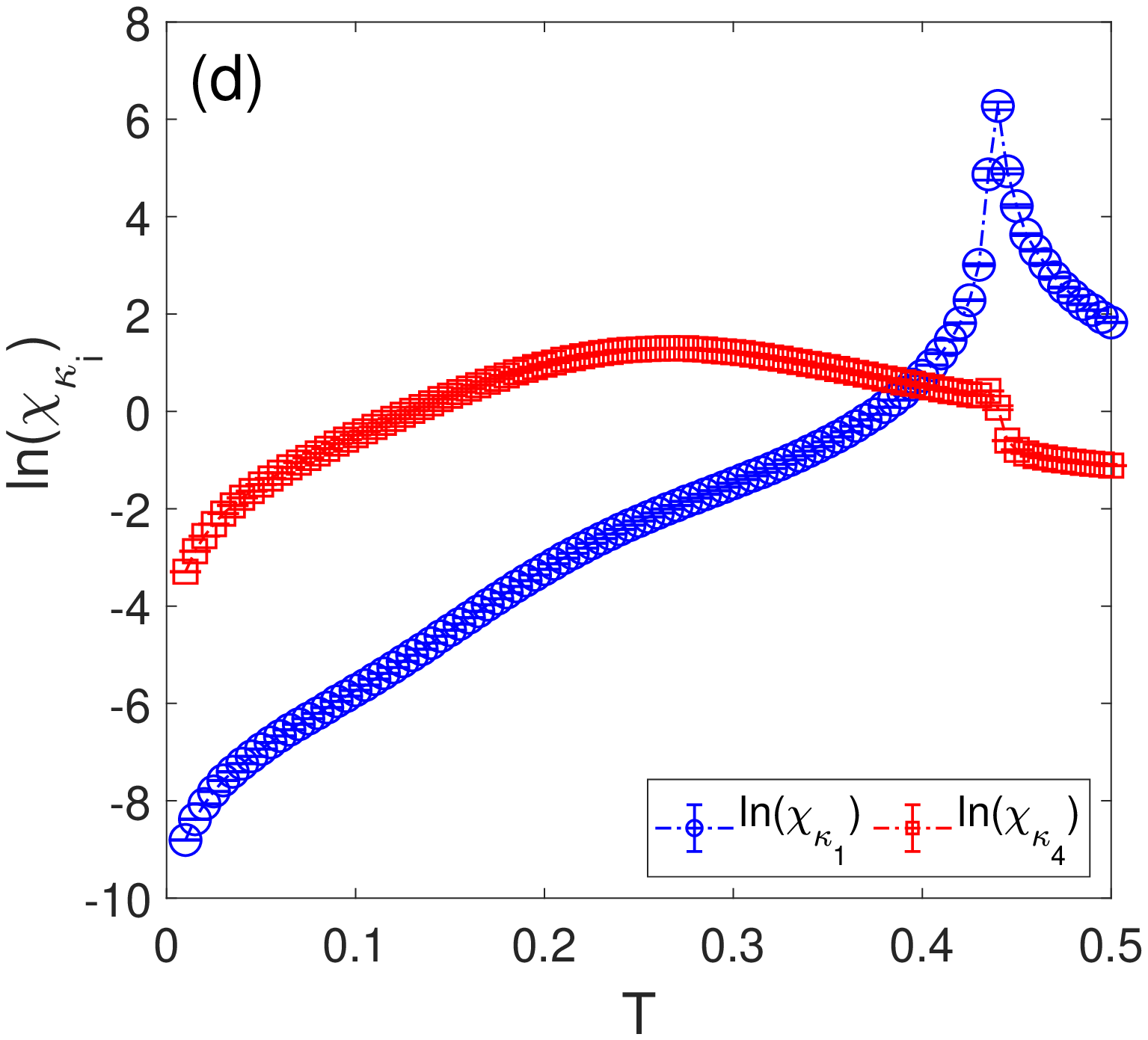}\label{fig:Chi_susc_q4_x08}}
\subfigure{\includegraphics[scale=0.34,clip]{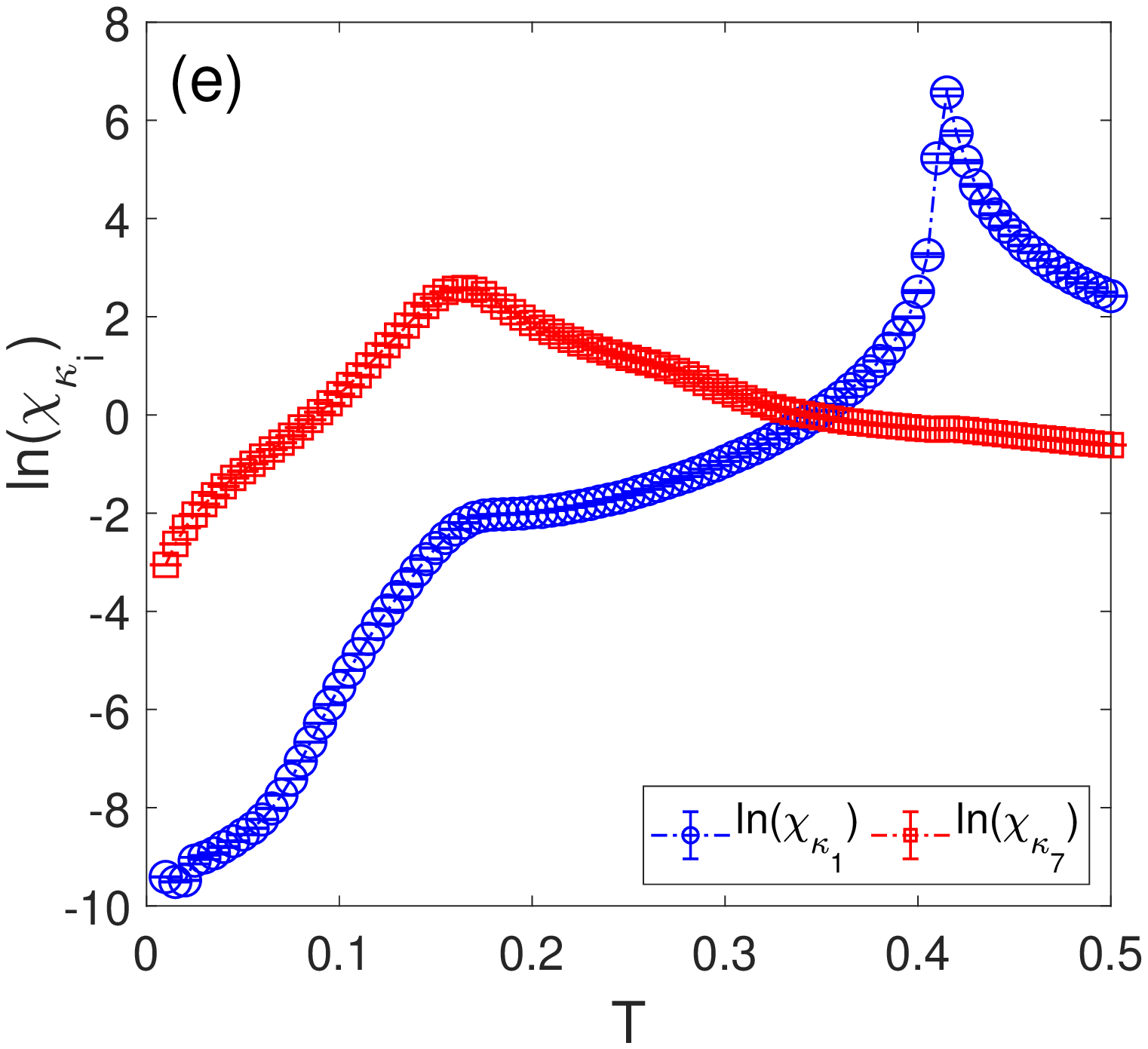}\label{fig:Chi_susc_q7_x08}}
\subfigure{\includegraphics[scale=0.34,clip]{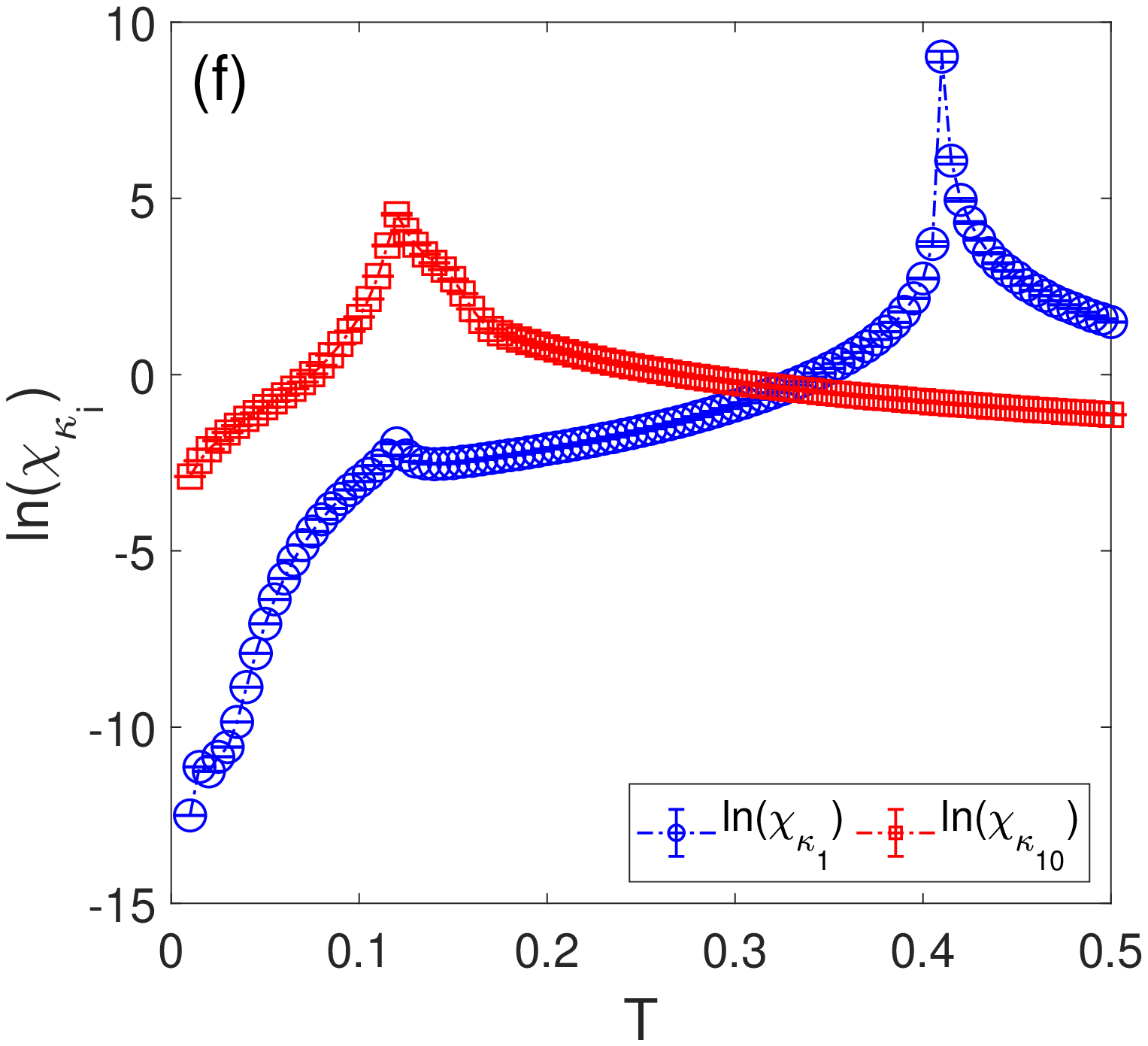}\label{fig:Chi_susc_q10_x08}}
\caption{Temperature dependencies of the standard and generalized chiral susceptibilities for $q=4$ (left column) , $q=7$ (middle column) and $q=10$ (right column), corresponding to $\Delta = 0.4$ (upper row) and $\Delta = 0.8$ (lower row).}
\label{fig:chir_susc}
\end{figure}

It is also interesting to notice that in the low-temperature AFM\textsubscript{1} phase the magnetic correlation function decays extremely slowly. The associated critical exponent $\eta_{m_1}$ (of the order of $10^{-4}$) is about two orders of magnitude smaller than within the AFM\textsubscript{0} phase. This is also reflected in even more dramatic drop of the magnetic susceptibility $\chi_{m_1}$ at the AFM\textsubscript{1}-AFM\textsubscript{0} phase transition, as shown in Fig.~\ref{fig:magn_susc}. This behavior can be ascribed to the suppressed magnetic fluctuations in the AFM\textsubscript{1} phase, as demonstrated in Fig.~\ref{fig:Absolute_angles}.  

Let us remind us that in addition to the magnetic and nematic orderings, in the present frustrated model there are also (generalized) chirality orderings in the system. In Figs.~\ref{fig:chir} and~\ref{fig:chir_susc} we present temperature dependencies of the standard and generalized staggered chiralities, $\kappa_1$ and $\kappa_q$, and the associated generalized chiral susceptibilities, $\chi_{\kappa_1}$ and $\chi_{\kappa_q}$. We note that in the previously studied $q = 3$ case both of the chiral order parameters vanished only close to the transition to the paramagnetic phase. For $q>3$, their behavior changes depending on $q$. In particular, for $0 < \Delta < 0.5$, $\kappa_1$ vanishes together with the magnetic order parameter $m_1$ at either the AFM\textsubscript{1}-ANq, or, for the values of $q$ where the AFM\textsubscript{2} phase exists, at the AFM\textsubscript{2}-ANq phase transition. Thus, only $\kappa_q$ remains finite in the ANq phases. For $0.5 < \Delta < 1.0$, both chiralities remain finite in the intermediate AFM\textsubscript{0} phase for $q$ up to 6. Note that in Fig.~\ref{fig:Chi_or_q4_x08} the parameter $\kappa_4$ shows some decline below the transition to the paramagnetic phase, nevertheless, its value remains finite up to the temperature at which $\kappa_1$ vanishes. However, starting from $q = 7$, only $\kappa_1$ remains finite while $\kappa_q$ copies the behavior of $m_q$ and vanishes at the AFM\textsubscript{1}-AFM\textsubscript{0} or AFM\textsubscript{2}-AFM\textsubscript{0} transition. None of the chiral order parameters vanish at the AFM\textsubscript{1}-AFM\textsubscript{2} transition and all of the phases display at least one form of chiral ordering for all values of $q$.

\begin{figure}[t!]
\centering
\subfigure{\includegraphics[scale=0.5,clip]{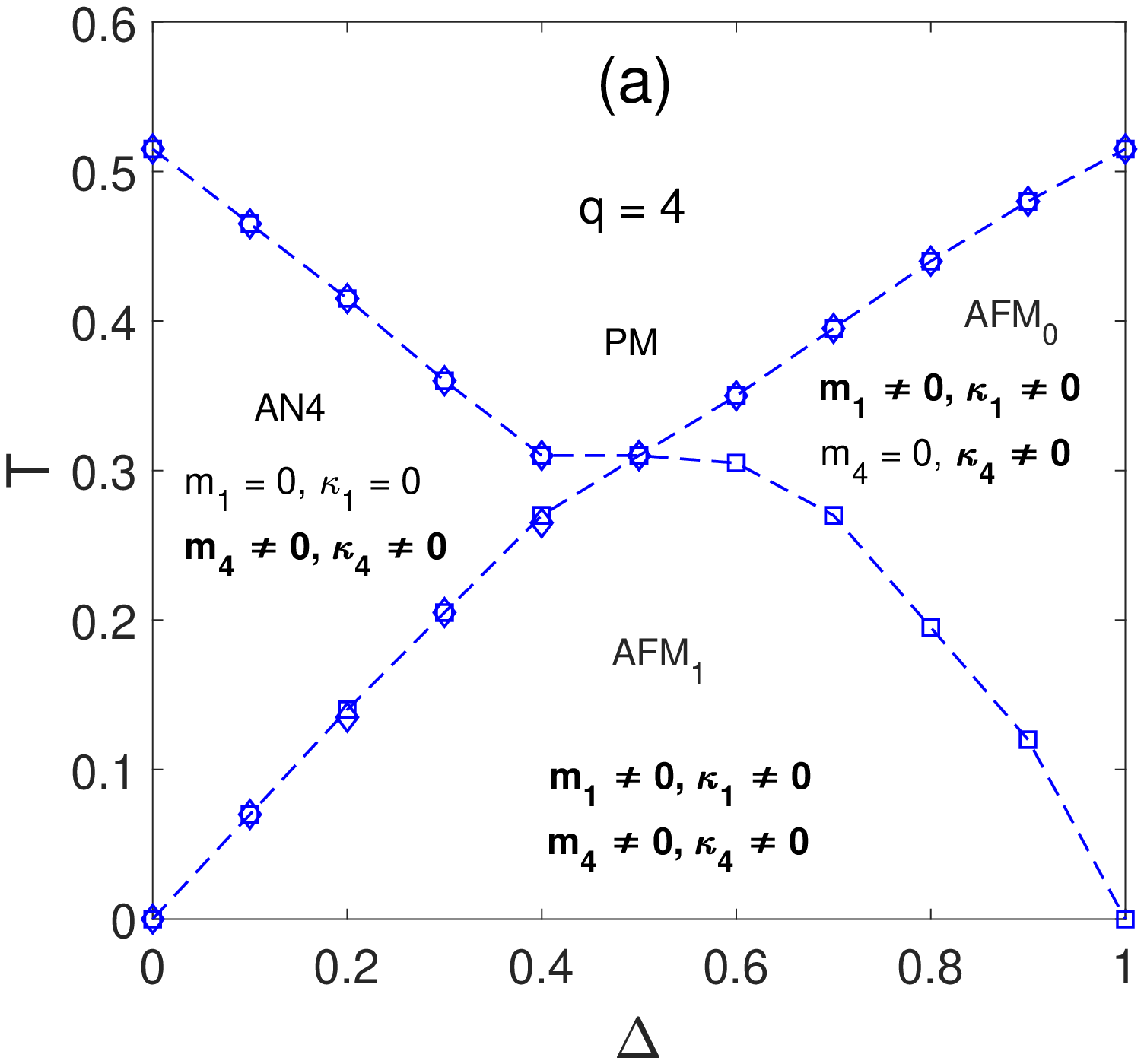}\label{fig:PD_q4}}
\subfigure{\includegraphics[scale=0.5,clip]{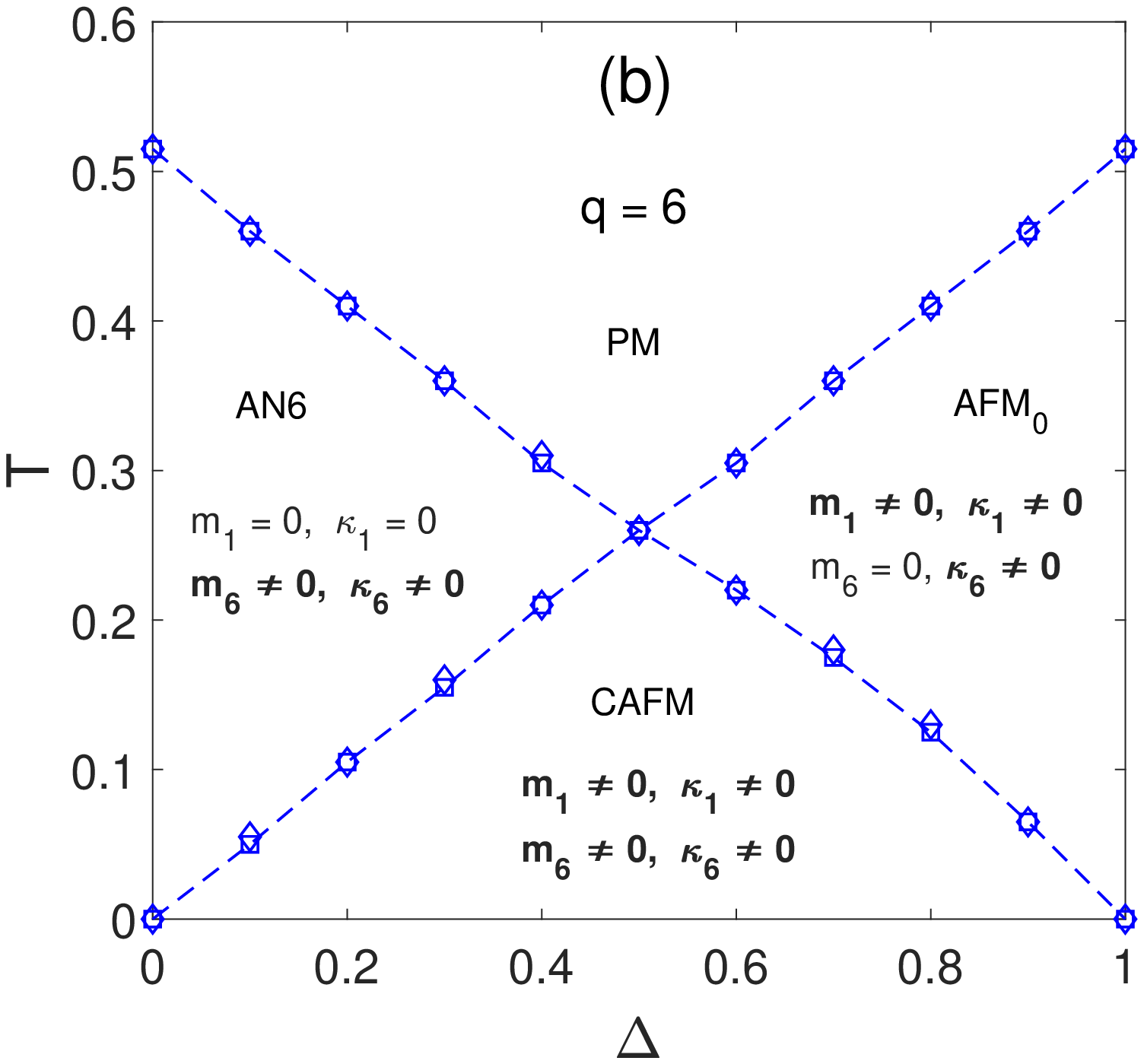}\label{fig:PD_q6}}\\
\subfigure{\includegraphics[scale=0.5,clip]{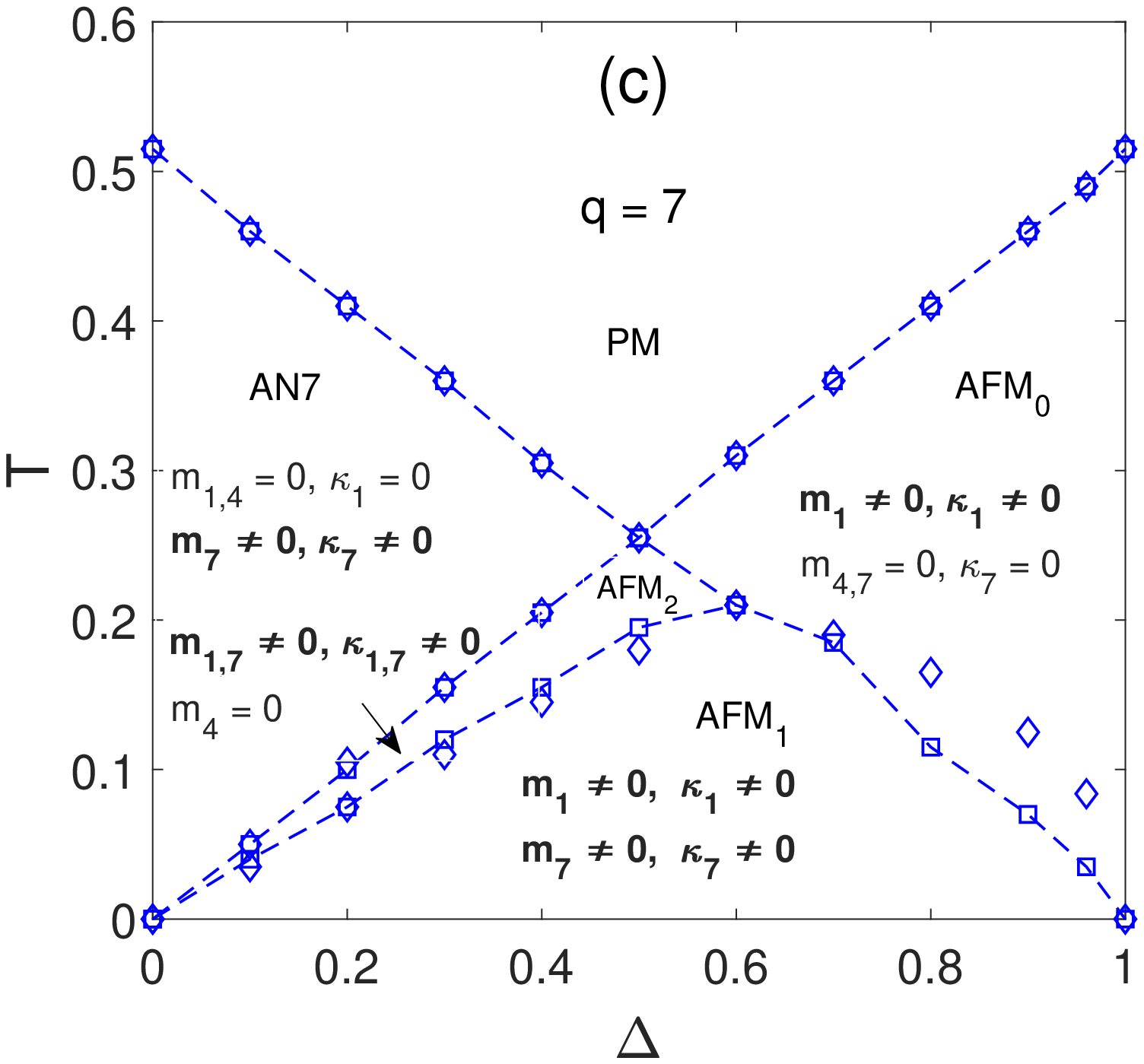}\label{fig:PD_q7}}
\subfigure{\includegraphics[scale=0.5,clip]{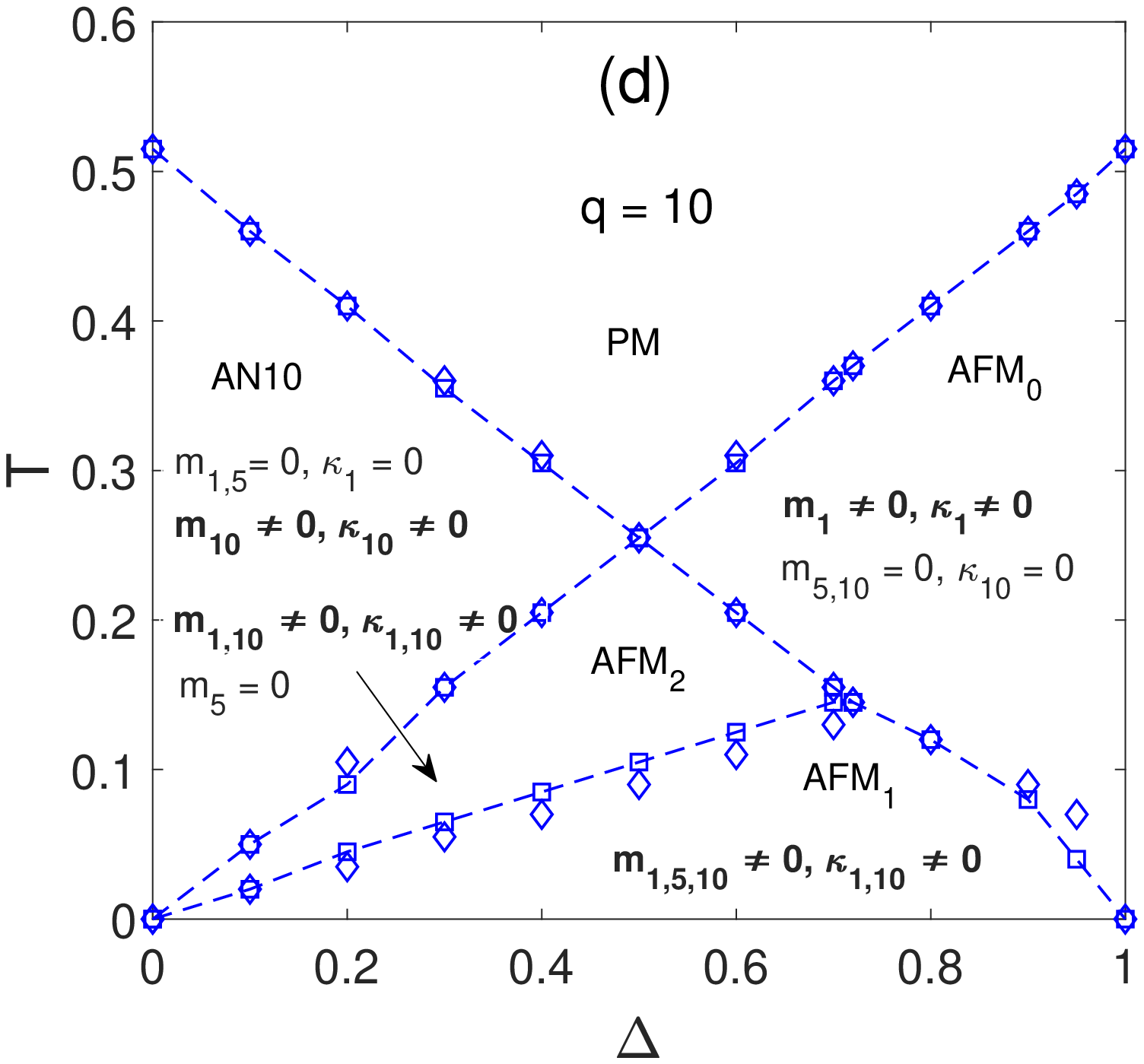}\label{fig:PD_q10}}
\caption{Phase diagrams in $\Delta-T$ parameter plane, for representative values of $q$. Diamond (square) symbols represent phase boundaries located from the peaks of the specific heat (generalized magnetic order susceptibility). In (a) the AFM\textsubscript{0}-AFM\textsubscript{1} phase boundary could not be estimated based on the specific heat due to the absence of a distinct peak for $q=4$. The captions display the observed phases and the corresponding order parameters, with those taking finite values in the respective phases highlighted in bold.}
\label{fig:PD}
\end{figure}

The results are summarized in the phase diagrams shown in Fig.~\ref{fig:PD}, which were constructed using the order parameter susceptibility peaks (squares) as well as the peaks in the specific heat measurements (diamonds). The phase boundaries roughly estimated by these two methods show a rather good correspondence, except for the AFM\textsubscript{1}-AFM\textsubscript{0} and AFM\textsubscript{1}-AFM\textsubscript{2} phase boundaries, for which the specific heat peaks predict respectively higher and lower transition temperatures than the corresponding susceptibilities peaks. We note that for $q = 4$ and $q=5$ with larger $\Delta$ the specific heat curves do not provide reliable estimate of the location of the AFM\textsubscript{1}-AFM\textsubscript{0} phase boundary, as instead of sharp peaks they only show either broad shoulders ($q = 4$) or round and broad maxima ($q = 5$). Nevertheless, increasing the nematic parameter $q$ from 4 to 5 does not seem to alter the system's critical behavior, while $q = 6$ converts the low temperature AFM\textsubscript{1} phase to the CAFM phase. The topology of the phase diagram for $q = 6$, as shown in Fig.~\ref{fig:PD_q6}, is similar to the $q = 3$ case from our previous work~\cite{Lach_Zukovic_2020}. In fact, all the values of $q$ divisible by 3 (up to $q = 15$ studied in this work) show the same magnetic phase diagram topology and no apparent deviations from the $q = 3$ case.

\begin{figure}[t!]
\centering
\includegraphics[scale=0.7,clip]{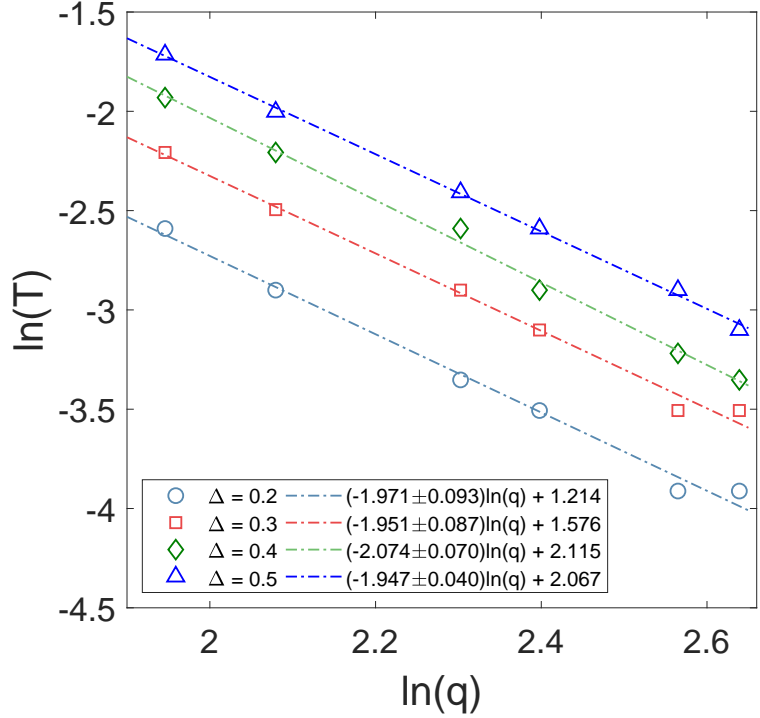}
\caption{Fits of the AFM\textsubscript{1} - AFM\textsubscript{2} transition temperature dependence on the nematic parameter $q$, for various $\Delta$.}
\label{fig:q-2_fit_linear}
\end{figure}

For $q = 7$ there is another change in the phase diagram topology. Namely, the AFM\textsubscript{1} - ANq phase transition line bifurcates, creating an additional AFM\textsubscript{2} phase. The region occupied by this new phase increases with $q$ at the cost of the AFM\textsubscript{1} phase. The AFM\textsubscript{1}-AFM\textsubscript{2} transition temperature appears to decrease with $q$ for all values of $\Delta$ as $q^{-2}$, as shown in Fig.~\ref{fig:q-2_fit_linear}. There are no further changes in the phase diagram topology for $q$ up to 14, the largest studied value of $q$ nondivisible by 3.

\section{Summary and discussion}
\hspace*{5mm} In this study, we have extended our previous work~\cite{Lach_Zukovic_2020} on the critical behavior of the generalized $XY$ model on a triangular lattice with AFM and generalized ANq interactions for $q=3$, by considering higher values of the generalized nematic parameter up to $q=15$. As previously shown \cite{zukovic_frustrated_2016}, the inclusion of the ANq interaction for $q = 3$ and 6 changes the ground state from the usual AFM structure with $2\pi/3$ relative phase angles to a peculiar canted (CAFM) state. This change is caused by the conflict of the AFM and ANq interactions on the geometrically frustrated lattice and leads to phase diagrams containing three QLRO phases for all values of $q$ divisible by 3 at least up to $q = 15$, the highest value studied in this work. Besides the CAFM phase with unsaturated values of the magnetic and generalized nematic order parameters down to zero temperatures, there are two intermediate-temperature phases with pure AFM and ANq orderings.

In the case of $q$ nondivisible by 3, the well known phase diagram topology for $q = 2$ changes first at $q = 4$ and then again at $q = 7$. In particular, for $q=4$ the AFM\textsubscript{0} phase with purely AFM correlations separates from the AFM\textsubscript{1} phase with the coexistent AFM and AN4 correlations. In the AFM\textsubscript{1} phase, both the magnetic and generalized nematic order parameters reach saturated values at low temperatures and the snapshots show typical AFM structure. Increasing the nematic parameter to $q=7$ leads to the appearance of the AFM\textsubscript{2} phase in a part of the region previously occupied by the AFM\textsubscript{1} phase. This new phase still shows both the AFM and ANq orderings, however, the typical AFM spin structure disappears. Instead, for each sublattice there are $\ceil{q/2}$ possible spin orientations with different weights belonging to the same half plane. The order parameter for the AFM\textsubscript{1}-AFM\textsubscript{2} phase transition is $m_{\ceil{q/2}}$.

All the observed phases display at least some kind of chiral LRO. In the low-temperature AFM\textsubscript{1} and AFM\textsubscript{2} phases, as well as the frustrated CAFM phase, both the standard $\kappa_1$ and the generalized $\kappa_q$ staggered chiralities remain finite. $\kappa_1$ vanishes at the transition to the ANq phase from the low-temperature phases for all $q$ so that inside the ANq phases only $\kappa_q$ remains finite. On the other hand, at the transition to the AFM\textsubscript{0} phase from the low-temperature phases, $\kappa_q$ remains nonzero for $q$ up to 6, while starting with $q = 7$ it drops to zero together with the nematic $m_q$ order parameter at the AFM\textsubscript{1}-AFM\textsubscript{0} or AFM\textsubscript{2}-AFM\textsubscript{0} transition and thus inside the AFM\textsubscript{0} phase only $\kappa_1$ remains finite. 

It is interesting to compare the results for the present frustrated AFM-ANq models with those obtained for the related nonfrustrated models with FM-Nq interactions on a square lattice~\cite{Poderoso-2011, Canova-2014, Canova-2016}. It is worth to note that for $q=2$ both systems display the same phase diagram topology (see e.g.,~\cite{Park-nematic,Hubscher_2013}). However, for $q>2$, in the former case the conflict between the AFM and ANq interactions caused by the geometrical frustration arises for $q$ divisible by 3, which is absent in the latter models, and leads to the formation of the CAFM phase. For $q$ nondivisible by 3, there is no conflict between the two interactions and the results can be more easily compared. In the FM-Nq models as $q$ increases a new QLRO phase appears for $q = 4$ at low temperatures due to the competition between the FM and N4 interactions, denoted in Ref.~\cite{Canova-2016} as F\textsubscript{1} phase. Similar behavior was observed in the present AFM-AN4 model, accompanied with the emergence of the AFM\textsubscript{1} phase, albeit there might be a crossover rather than a standard phase transition to this phase. Further increase of the nematic parameter $q$ to 5 results in another change in the phase diagram topology of the FM-N5 model, featuring the F\textsubscript{2} phase, while the topology of the AFM-AN5 model seems to remain unchanged. Nevertheless, the AFM\textsubscript{2} phase, which might be viewed as a counterpart of the F\textsubscript{2} phase, appears in the present models for $q=7$. Thus, except for the CAFM phase in the AFM-ANq models, which has no analog in the nonfrustrated FM-Nq counterparts, the nature of the remaining phases in the two cases can be related. Namely, the respective types of orderings in the square-lattice FM-Nq models can be observed on each of the three sublattices of the triangular-lattice AFM-ANq models. We think that because of the increased ``stiffness'' of the spin distributions in the latter models, due to the AFM constraints between spins belonging to different sublattices, a larger value of $q$ is required for splitting the unimodal distribution in the AFM\textsubscript{1} phase to facilitate the emergence of the multimodal distribution in the AFM\textsubscript{2} phase. No further topology changes are observed in either models with the increasing $q$, nevertheless, the area occupied by the F\textsubscript{2} and AFM\textsubscript{2} phases increases due to the power-law decrease of the F\textsubscript{2}-F\textsubscript{1} and AFM\textsubscript{2}-AFM\textsubscript{1} transition temperatures as $q^{-2}$.

Finally, the goal of the present study was the evolution of the phase diagram topology of the model with the increasing higher-order coupling and we have not attempted to determine the character of all the identified phase transitions. As we found out when performing such an analysis for the $q=3$ case~\cite{Lach_Zukovic_2020}, this task for the present frustrated systems in such a broad parameter space would require enormous amount of additional simulations and thus we leave it for future considerations. Nevertheless, the interesting results obtained for their nonfrustrated counterparts~\cite{Poderoso-2011, Canova-2014, Canova-2016}, featuring phase transitions belonging to a variety of universality classes, would suggest that it is well worth trying.

\begin{acknowledgments}
This work was supported by the Scientific Grant Agency of Ministry of Education of Slovak Republic (Grant No. 1/0531/19) and the Slovak Research and Development Agency (Contract No. APVV-18-0197). The authors would also like to thank the Joint Institute for Nuclear Research in Dubna, Russian Federation for the use of their Govorun Supercomputer.
\end{acknowledgments}


\end{document}